\documentclass{aastex701}

\begin{document}


\title{A Quantitative Method to Investigate the Spectral Class Time Evolution of Novae}

\author[0000-0002-3563-819X]{M. Kopsacheili}\thanks{SO Incoming Visitor}
\affiliation{Institute of Space Sciences (ICE, CSIC), Barcelona, Spain}
\affiliation{Institut d’Estudis Espacials de Catalunya (IEEC),  Barcelona, Spain}
\email[show]{kopsacheili@ice.csic.es}  

\author[orcid=0000-0002-7759-106X]{M. Guerrero} 
\affiliation{Instituto de Astrofísica de Andalucía–CSIC, Granada, Spain}
\email{mar@iaa.es}

\author[0009-0008-2354-0049]{D.A. Vasquez-Torres}
\affiliation{Instituto de Radioastronomía y Astrofísica (IRyA), Universidad Nacional Autónoma de México (UNAM), Morelia, Mexico}
\email{d.vasquez@irya.unam.mx}


\begin{abstract}

Novae are spectroscopically classified as \ion{Fe}{2} or He/N according to the dominance of \ion{Fe}{2}, or He and N emission lines in their pre-nebular spectra, while spectra exhibiting both are classified as hybrid. Whereas the spectral class of a nova is known to correlate with its progenitor properties, a recent detailed spectroscopic follow-up of ten novae suggests instead a time evolution from ``early He/N'' to ``Fe II'' and to ``late He/N''. 
We investigate here this apparent contradiction using a spectroscopic dataset of 26 novae from the Astronomical Ring for Amateur Spectroscopy (ARAS) Database. 
For this purpose we introduce a quantitative coefficient, $\kappa$, defined as the ratio of the flux of the \ion{Fe}{2} $\lambda5169$ line to the mean flux of five He and N emission lines, with $\kappa>1$ indicating \ion{Fe}{2}-dominated spectra and $\kappa<1$ indicating He/N-dominated spectra. This coefficient is shown to provide a reliable
quantitative tool for tracing nova spectral evolution while reducing biases from incomplete coverage. 
A significant fraction of the novae in our sample exhibit at least one transition between \ion{Fe}{2}- and He/N-dominated phases. 
The time evolution of $\kappa$ closely follows their optical light-curves, repeatedly crossing unity in novae exhibiting jitters, which reveal non-monotonic spectral evolution. 
Since the light-curve shapes and their time-scales depend on intrinsic progenitor properties, particularly white dwarf mass, they also determine the duration of \ion{Fe}{2}- and He/N-dominated phases, introducing a bias in spectral classification when it relies on early, sparse spectra.

\end{abstract}



\section{Introduction}
Nova eruptions occur on the surface of white dwarfs (WDs) accreting material in close binary systems from a main-sequence (classical nova, CN) or a giant (symbiotic nova) companion \citep{1976MNRAS.176...53G,2008clno.book.....B}.  
Material is typically transferred from the companion to an accretion disk and eventually onto a thin layer on the WD surface. 
As material builds up, the pressure and temperature at the base of this layer increase until critical values are reached for nuclear fusion to be ignited under degenerate conditions, leading to a thermonuclear runaway (TNR).

The TNR rapidly converts hydrogen into heavier elements and releases large amounts of energy.  
This produces a luminous but transient nova outburst, in which the brightness can increase up to 15 magnitudes \citep{1957gano.book.....G}. 
The energy released in the TNR expels the majority of the unburnt hydrogen from the WD's surface as a fast expanding shell, with velocities reaching up to $\rm \sim 1500\,km\,s^{-1}$.   
The ejecta disrupts the accretion disk and accretion is halted. 
With time, accretion resumes, a new accretion disk forms, and eventually another nova outburst will occur. 
When the recurrence time is short enough that at least two outbursts have been detected, the nova is said to be recurrent.

The TNR and the interaction of the ejecta with the circumstellar medium are complex processes, resulting in a large variety of nova observational properties.  
The brightness of a nova rises sharply until its peak in a few days (1-3 d), but the shape of the light-curve and decline time-scales afterwards vary largely among novae. 
These differences have motivated their classification into different light-curve shapes \citep{2010AJ....140...34S} and speed class \citep{1957gano.book.....G}. 
Spectra also vary largely among novae. 
The most prominent spectral features are the Balmer series, low-excitation \ion{Fe}{2} lines, and high-excitation lines of He and N, such as \ion{He}{1}, \ion{He}{2}, \ion{N}{2}, and \ion{N}{3}. 
The relative dominance of these features near peak brightness has traditionally been used to classify novae into “\ion{Fe}{2}” or “He/N” types \citep{1992AJ....104..725W,2012AJ....144...98W}. 
He/N novae usually present broad lines with very few or no absorption features, which can be attributed to an optically thin gas expanding at high velocities. 
On the other hand, \ion{Fe}{2} novae have lower ejecta velocities, narrower lines and often they present prominent P Cygni absorption features indicating an optically thick expanding gas \citep{1992AJ....104..725W,2012AJ....144...98W}.

The appearance of these spectroscopic features is attributed to different gas conditions and/or the emission arising from different regions. \citet{2012AJ....144...98W} suggested that the \ion{Fe}{2} emission originates in the envelope of gas produced by the secondary star in the binary system, given the similarity of these spectral features to those of late-type stars and stellar winds. On the other hand, the He and N emission would originate from the WD itself, particularly in light of the high-velocity spectroscopic features. \citet{2012BASI...40..185S,2013A&A...559L...7S} support that the \ion{Fe}{2} or He/N emission is related to different phases of the gas, linked to its optical thickness. When the ejecta is optically thick, the spectrum is dominated by \ion{Fe}{2} lines, as the so-called {\it iron curtain} absorbs the high-energy radiation and prevents seeing the inner, hotter regions. As the ejecta expands and becomes optically thin, this curtain lifts, allowing higher-energy photons to escape, revealing emission from more highly ionized species such as He and N.
A small fraction of novae exhibits a transition between the “\ion{Fe}{2}” and “He/N” class and are therefore dubbed as ``hybrid'' or ``\ion{Fe}{2} b'' class.

The evolutionary properties of novae, including their decline rate and brightness, and their spectroscopic class have been linked with the properties of the progenitor systems, most importantly WD mass, but also metallicity and presence of magnetic field. 
Then more massive WD would produce bright, fast novae preferentially found in the Galactic Plane \citep{1992A&A...266..232D,1994A&A...287..403D} that are predominantly classified as He/N novae, whereas fainter, slow novae are found at higher Galactic latitudes and predominantly classified as \ion{Fe}{2}
novae \citep{1998ApJ...506..818D}. 
These early findings are supported by a recent investigation of 46 Galactic novae that conclude that fainter slow novae at high Galactic latitudes are consistent with an older stellar population \citep{2025ApJ...981..198C}.

Similar trends are reported for extragalactic nova populations. 
Brighter, fast novae are preferentially located in the galactic plane of M\,31, while slower and fainter novae are more commonly found in its bulge \citep{2011ApJ...734...12S}. 
The higher occurrence of fast novae, including He/N and hybrid novae, in the Large Magellanic Cloud (LMC) compared to M\,31 \citep{2013AJ....145..117S} suggests that the higher fraction of fast novae, including He/N and \ion{Fe}{2} b novae, in the LMC compared to M\,31 may be related to differences in their underlying stellar populations. 
Thus the younger stellar population of the LMC produces binary progenitors hosting more massive WDs than those in M31. 
This interpretation is further supported by \citet{2026ApJS..283...24S}, who showed that He/N and \ion{Fe}{2} b novae occupy the high-mass end of the WD mass distribution.


These well-established results seems somehow in contradiction with a possible novae spectral evolution. 
\citet{2012AJ....144...98W} suggested that most novae are hybrid and undergo a transition from the \ion{Fe}{2} to the He/N class, but such transition is often missed observationally due to the short duration of the different phases. 
Recently, \citet{2024MNRAS.527.9303A} presented a throughout spectral and photometric follow-up of ten novae and determined their spectral class at different epochs. 
They concluded that all of them pass through different spectral phases: 
(i) an “early He/N” phase, 
(ii) an “\ion{Fe}{2}” phase, and 
(iii) a “late He/N” phase.  
This further supports the idea that evolution from one class to another occurs in most, if not all, novae, which can be attributed to distinct gas components or driven by changes in the ejecta physical conditions affecting its ionization and opacity \citep{2012AJ....144...98W,2012BASI...40..185S,2014ASPC..490..145S}.

The availability of multi-epoch spectra for a substantial number of novae in the Astronomical Ring for Amateur Spectroscopy (ARAS) Database \citep[hereafter ARAS,][]{2019CoSka..49..217T} provides the opportunity to extend to a larger sample of novae the investigation of their spectral evolution and to assess its link with their evolutionary timescales and the properties of their progenitor systems. 
We have selected spectra for 26 novae to investigate their spectral evolution before entering the nebular phase.  
A coefficient defined as the ratio of the flux of the \ion{Fe}{2} 5169\AA\ line to the mean flux of the \ion{He}{1} 6678\AA\ and 7065\AA, 
and \ion{N}{2} 4638\AA, 5679\AA, and 5755\AA\ lines has been introduced to assess quantitatively the spectral evolution of novae.  The time variation of this coefficient is compared to the photometric evolution to assess the link between photometric and spectroscopic phases in the nova evolution.

The structure of the paper is as follows.  
Sect.~2 introduces and describes the spectroscopic and photometric datasets used in this paper. 
Sect.~3 describes the spectral classification and introduces the spectral evolution coefficient.  
The results are presented in Sect.~4 and the discussion in Sect.~5.  
Sect.~6 presents the concluding remarks of this work.


\section{Data}

\subsection{Spectra from the ARAS Database of Novae}

ARAS\footnote{\url{https://aras-database.github.io/database/index.html}} is a spectral database built by citizen astronomers.  
It includes spectroscopic observations of multiple types of variable and eruptive stars such as novae.   
The ARAS Database was search for spectroscopic observations of novae available as of 2026 March 1.

The ARAS Database includes 1014 multi-epoch spectra of 77 novae. 
This dataset was reduced to the 739 flux-calibrated spectra of 40 novae, and further limited to a sample of 26 objects for which there are at least four calibrated spectra, so as to allow us investigating their spectral time evolution. 
The list of those 26 novae is provided in Table ~\ref{tab:novae}, together with the number of available spectra, their eruption date (when available), the time in days required for the nova to fade by 2 ($t_2$) and 3 ($t_3$) magnitudes from maximum light, the speed class defined by the value of $t_2$ \citep{1957gano.book.....G}, the light-curve shape, and the spectral class identified in this work.

\begin{table*}
\centering
\begin{tabular}{lrcrrccc}
\hline
Nova Name & 
\multicolumn{1}{c}{Spectra} & 
\multicolumn{1}{c}{$t_\mathrm{max}$} & 
\multicolumn{1}{c}{$t_2$} & 
\multicolumn{1}{c}{$t_3$} & 
\multicolumn{3}{c}{\underline{~~~~~~~~~~~~~~~~~~~~~~Nova Classification~~~~~~~~~~~~~~~~~~~~~~}} \\
 & 
\multicolumn{1}{c}{\#} & 
\multicolumn{3}{c}{} & 
\multicolumn{1}{c}{Speed} & 
\multicolumn{1}{c}{Light-curve} & 
\multicolumn{1}{c}{Spectral} \\
\multicolumn{5}{c}{} & 
\multicolumn{1}{c}{Class} & 
\multicolumn{1}{c}{Shape} & 
\multicolumn{1}{c}{Class} \\
[-0.66cm]
\multicolumn{2}{c}{} & 
\multicolumn{1}{c}{(MJD)} & 
\multicolumn{1}{c}{(d)} & 
\multicolumn{1}{c}{(d)} & 
\multicolumn{3}{c}{}\\[0.3cm]
\hline
\multicolumn{8}{c}{Classical novae with spectral class transitions} \\
\hline
Nova Cen\,2013 (V1369 Cen) & 45 & 56628.69 &  37 & 65 & moderately fast & D & \ion{Fe}{2} $\rightarrow$ He/N \\ 
Nova Del\,2013 (V339 Del) & 52 & 56518.58 & 14 & 29 & fast & PP & \ion{Fe}{2} $\rightarrow$ He/N\\ 
Nova Oph 2015 (V2944 Oph) & 5 & 57152.00 & 16 & 145 & fast & J & \ion{Fe}{2} $\rightarrow$ He/N\\
Nova Sgr\,2015b (V5668 Sgr) & 8 &  57120.08 & 73 & 78 & moderately fast & D & \ion{Fe}{2} $\rightarrow$ He/N\\
Nova Sco\,2016 (V1655 Sco) & 10 & 57549.63 & 39 & 50 & moderately fast & D & \ion{Fe}{2} $\rightarrow$ He/N\\
Nova Cen\,2017 (V1405 Cen) & 11 & 57890.28 & 83 & 115 & slow & J & \ion{Fe}{2} $\rightarrow$ He/N\\ 
Nova Lup\,2018 (V408 Lup) & 5 & 58272.43 & 42 & 87 & slow & J & \ion{Fe}{2} $\rightarrow$ He/N\\ 
Nova SMC\,2019 & 9 & 58667.28 & $\dots$ & $\dots$ & $\dots$ & $^\star$S & \ion{Fe}{2} $\rightarrow$ He/N\\ 
Nova Per\,2020 (V1112 Per) & 60 & 59178.81 & 26 & 34 & moderately fast & $^\star$D & \ion{Fe}{2} $\rightarrow$ He/N\\ 
Nova Vul\,2024 (V615 Vul) & 35 & 60521.83 & $^\star$5 & $^\star$9 & very fast & $^\star$O & \ion{Fe}{2} $\rightarrow$ He/N\\
Nova SMC\,2020 & 4 & $\dots$ & $\dots$ & $\dots$ & $\dots$ & $^\star$S & \ion{Fe}{2} $\rightarrow$ He/N\\
Nova Cyg\,2014 (V2659 Cyg) & 32 & 56646.79 & 115 & 140 & slow & J & \ion{Fe}{2} $\leftrightarrows$ He/N\\
Nova Cas\,2020 (V1391 Cas) & 88 & 59076.05 & 110 & 120 & slow & D & \ion{Fe}{2} $\leftrightarrows$ He/N\\
Nova Cas\,2021 (V1405 Cas) & 143 & 59291.42 & 166 & 175 & very slow & J & He/N $\leftrightarrows$ \ion{Fe}{2} \\ 
Nova Vul\,2021 (V606 Vul) & 51 &  59411.49 & $\dots$ & 96 & moderately fast~/~slow & P & He/N $\leftrightarrows$ \ion{Fe}{2} \\ 
\hline
\multicolumn{8}{c}{Classical novae without spectral class transition, but a trend on $\kappa$} \\
\hline
Nova Sco\,2013 (V1533 Sco) & 7 & $^{**}$56447.50 & 8 & $^\star$17 & fast & $^\star$D & \ion{Fe}{2} \\
Nova Sgr\,2016b (V5853 Sgr) & 4 & 57608.53 & 34 & 63 & moderately fast & S & \ion{Fe}{2} \\ 
Nova Sgr\,2016c (V5855 Sgr) & 6 & 57681.38 & 12 & 19 & fast & J & \ion{Fe}{2} \\ 
Nova Sgr\,2020d (V6593 Sgr) & 6 & 59124.98 & $^\star$18 & $^\star$33 & $^\star$fast & $^\star$S & \ion{Fe}{2} \\ 
Nova Sgr\,2021a (V6594 Sgr) & 4 & 59299.76 & 29 & 38 & moderately fast & D & \ion{Fe}{2} \\ 
Nova Her\,2021 (V1674 Her) & 25 & 59377.55 & 1 & 2 & very fast & S & He/N \\ 
\hline
\multicolumn{8}{c}{Novae without spectral class transition and unclear trend on $\kappa$} \\
\hline
Nova Sct\,2018 (V613 Sct) & 9 & 58298.58  & 42 & 66 & moderately fast & S & \ion{Fe}{2}\\ 
CN\,Cha & 4 & $\dots$ & $\dots$ & $\dots$ & very slow & $^\star$C & He/N $\rightarrow$ He/N ? \\
Nova Ser\,2025 (V0691 Ser) & 4 & 60874.91 & $^\star$12 & $^\star$23 & $^\star$fast & $^\star$S & \ion{Fe}{2}\\ 
\hline
\multicolumn{8}{c}{Recurrent novae} \\
\hline
U\,Sco & 27 & $^\star$59737.27 & $^\star$1 & $^\star$4 & $^\star$very fast & PP & He/N\\
RS\,Oph & 60 & $^\star$59435.25 & $^\star$5 & $^\star$9 & $^\star$very fast & P & He/N $\rightarrow$ \ion{Fe}{2} $\rightarrow$ He/N \\

\hline
\end{tabular}

\caption{
Sample of novae with suitable multi-epoch, flux-calibrated ARAS spectra. 
The sample is split into four different groups for classical and symbiotic novae (i) with spectral class transition, (ii) without spectral class transition with a trend on the value of $\kappa$ and (iii) without a clear trend on the value of $\kappa$, and (iv) recurrent novae.  The values of $t_{max}$ have been taken from \href{https://www.astronomerstelegram.org/}{The Astronomer's Telegrams}, while those of  $t_2$ and $t_3$ from the compilation by \citet{2025ApJ...993..232S}, apart from the value marked by the sign $^{**}$ which has been obtained by \citet{2025MNRAS.541..980C}. 
According to the values of $t_2$, the novae are then assigned the speed class 'very fast' ($t_2 < 10~\mathrm{d}$), 'fast' ($11~\mathrm{d} < t_2 < 25~\mathrm{d}$), 'moderately fast' ($26~\mathrm{d} < t_2 < 80~\mathrm{d}$), 'slow' ($81~\mathrm{d} < t_2 < 150~\mathrm{d}$), and 'very slow' ($150~\mathrm{d} < t_2$). 
The light-curve shape is also adopted from the compilation by \citet{2025ApJ...993..232S}. 
It follows the scheme presented by \citet{2010AJ....140...34S}, including S (smooth), P (plateau), PP (double plateau), D (dust dips), C (cusps), O (oscillations), F (flat-topped), and J (jitters) light-curve shapes. 
When any value has been determined by us, it is marked by the sign $^\star$. 
The last column lists the spectral class of the novae and their transitions, if registered in the present data.  
We note that the spectrum of U\,Sco has been reported to have evolved from an early He/N phase to an \ion{Fe}{2} phase \citep{2024MNRAS.527.9303A}, but this transition is missed in the spectra of the ARAS Database used here. 
}

\label{tab:novae}
\end{table*}

\subsection{Light-curves from the AAVSO}

To further assess the possible correlation between spectral and photometric variations, optical photometry of the target sample was retrieved from the American Association of Variable Stars Observers (AAVSO\footnote{\url{https://www.aavso.org}}; \citealt{2018AAS...23222306H}). 
The $B$, $V$, $Visual$, and $R$ photometric measurements were considered. 



\section{Spectral Classification}

\subsection{Data analysis} 
\label{sec:cont}

The main goal of this work is to assess the possible time evolution of the spectral classification of novae using multi-epoch optical spectra available in the ARAS Database and to search for links with the photometric evolution of their light-curves.  
The spectral classification of a classical nova is determined by the prevalence of \ion{Fe}{2} or He and N emission lines.  
The analysis will then focus on the following emission lines: 
\ion{Fe}{2} $\lambda$5169, \ion{He}{1} $\lambda\lambda$6678,7065, and \ion{N}{2} $\lambda\lambda$4638,5679,5755.


Given the large number of spectra, an optimization procedure for measuring emission-line fluxes has been designed and applied to each spectrum. 
This procedure includes the continuum determination and subtraction, measurement of the flux of a list of selected emission lines, and estimate of the signal-to-noise (S/N) ratio. 
Additional checks for the possible contribution of blended emission lines were applied. 
The full procedure is described in length in Appendix~\ref{ap.1}.

The final outcome for each spectrum is the flux measurement, S/N ratio, and a flag quality for blended lines of the \ion{Fe}{2} $\lambda$5169, \ion{He}{1} $\lambda\lambda$6678,7065, and \ion{N}{2} $\lambda\lambda$4638,5679,5755 emission lines. 
Ohter emission lines of \ion{Fe}{2}, He and N were excluded, as they are often blended with nearby spectral features. 
Furthermore only emission lines with positive fluxes, a S/N greater than 10, and not affected by blends with nearby lines were considered for the subsequent analysis.

\subsection{$\kappa$, a quantitative estimator of the nova spectral type}

The nova spectral class is assigned according to the prevalence of \ion{Fe}{2} or He and N emission lines. 
Such a ``qualitative'' criterion cannot be efficiently applied to large spectral datasets, as investigated here. 
They require applying a ``quantitative'' criterion to describe objectively the nova class.  
Therefore we introduce the coefficient $\kappa$, 
defined as the ratio between the flux of the \ion{Fe}{2} emission line at 5169~\AA\ and the mean flux of the \ion{He}{1} $\lambda\lambda$6678,7065 and \ion{N}{2} $\lambda\lambda$4638,5679,5755 emission lines (Eq.\ref{eq:k}). 
The mean flux is defined as the arithmetic mean, i.e. the sum of the fluxes of the lines detected divided by their number, regardless of whether individual lines are detected in emission. Hence, 
\begin{equation}
\kappa =
\frac{F(\mathrm{Fe\,II}\,\lambda5169)}
{\frac{1}{5}\sum_{i=1}^{5} F_i(\mathrm{He/N})}
=
\frac{\kappa_{\rm Fe}}{\kappa_{\rm He/N}},
\label{eq:k}
\end{equation}

where $F_i(\mathrm{He/N})$ are the fluxes of the
\ion{He}{1} $\lambda\lambda6678,7065$ and
\ion{N}{2} $\lambda\lambda4638,5679,5755$ emission lines.

For practical purposes, the numerator and denominator of the coefficient $\kappa$ are normalized by the corresponding H$\alpha$ flux:

\begin{equation}
\kappa_{\rm Fe} =
\frac{F(\mathrm{Fe\,II}\,\lambda5169)}
{F(\mathrm{H}\alpha)},
\qquad
\kappa_{\rm He/N} =
\frac{\frac{1}{5}\sum_{i=1}^{5} F_i(\mathrm{He/N})}
{F(\mathrm{H}\alpha)}.
\end{equation}

In this work, it will be adopted that spectra with values of $\kappa$ greater than unity ($\kappa_\mathrm{FeII} > \kappa_\mathrm{He/N}$) are considered to be \ion{Fe}{2}-dominated, whereas a value of $\kappa < 1$ ($\kappa_\mathrm{FeII} < \kappa_\mathrm{He/N}$) implies the opposite, i.e. a He/N-dominated spectrum.

\section{Results}

The time evolution of the spectral class of a nova as determined by the coefficient $\kappa$ is investigated in Figures~\ref{fig:allfigs}, \ref{fig:jitters}, \ref{fig:tendency}, and \ref{fig:rest}. 
For each nova with at least four suitable ARAS spectra, the left panel shows the extended light-curve in the $V$ band obtained from the AAVSO database, with vertical dashed lines marking the epoch of the spectra. The right panels zoom in the time intervals with spectral information, with the top panel displaying the $V$ light-curve, the middle panel the values of $\kappa_\mathrm{Fe}$ (blue) and $\kappa_\mathrm{He/N}$ (red), and the bottom panel the values of $\kappa$.

An inspection of these plots shows different behaviors of the $\kappa$ coefficient.  
These are organized into different categories, as listed in Table~\ref{tab:novae}, and described in the next subsections.

\subsection{Classical novae with transitions between spectral classes}

The top block in Table~\ref{tab:novae} includes novae that show spectra both dominated by \ion{Fe}{2} and by N and He lines. 
These can be split into two different groups, depending on whether they present a single or multiple transitions between spectral classes.

\subsubsection{Single transition from \ion{Fe}{2} to He/N: \ion{Fe}{2} $\rightarrow$ He/N. }

\begin{figure*}
\centering

\begin{minipage}{0.46\textwidth}
    \centering
    \includegraphics[width=\linewidth]{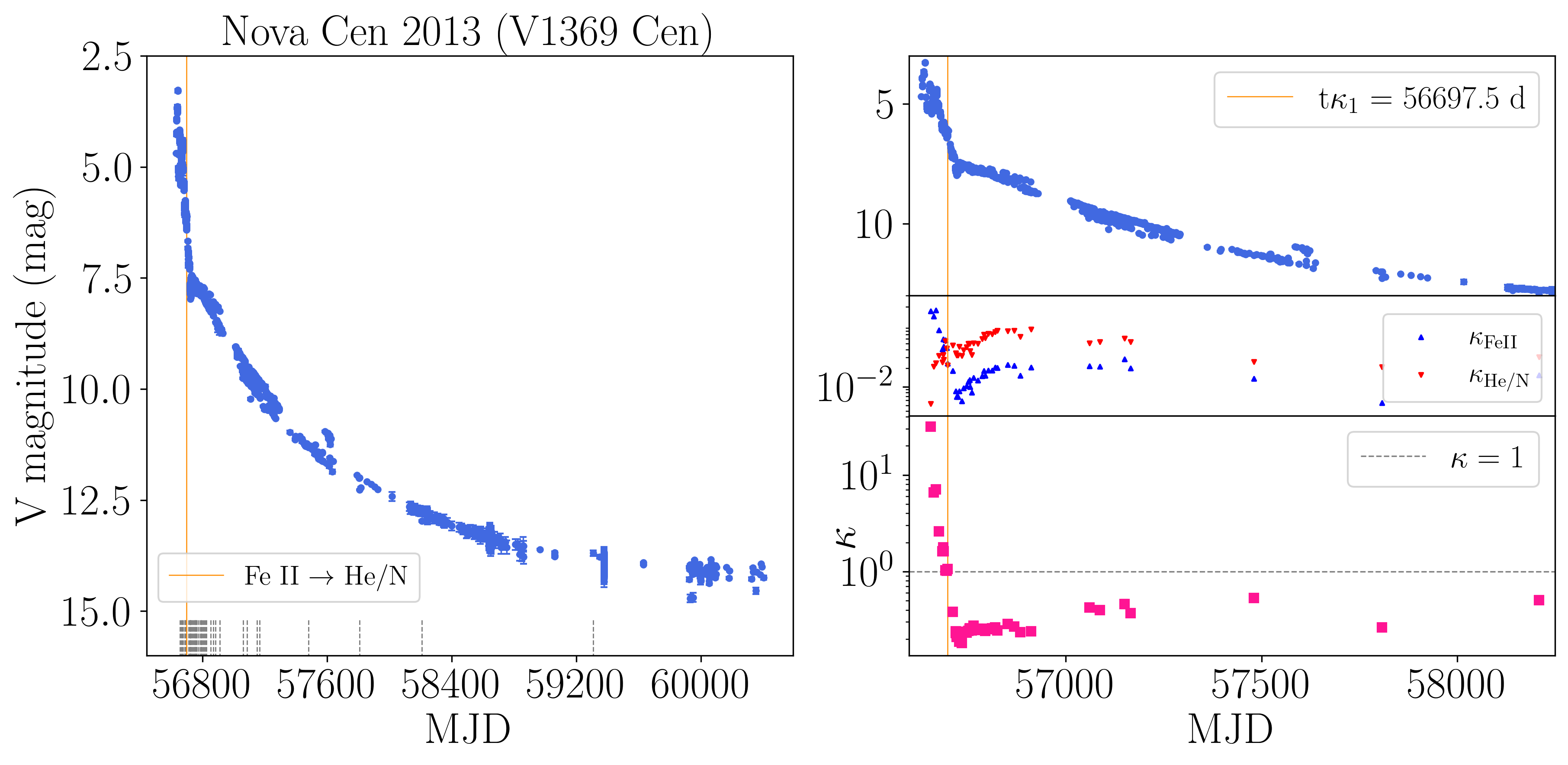}
    \includegraphics[width=\linewidth]{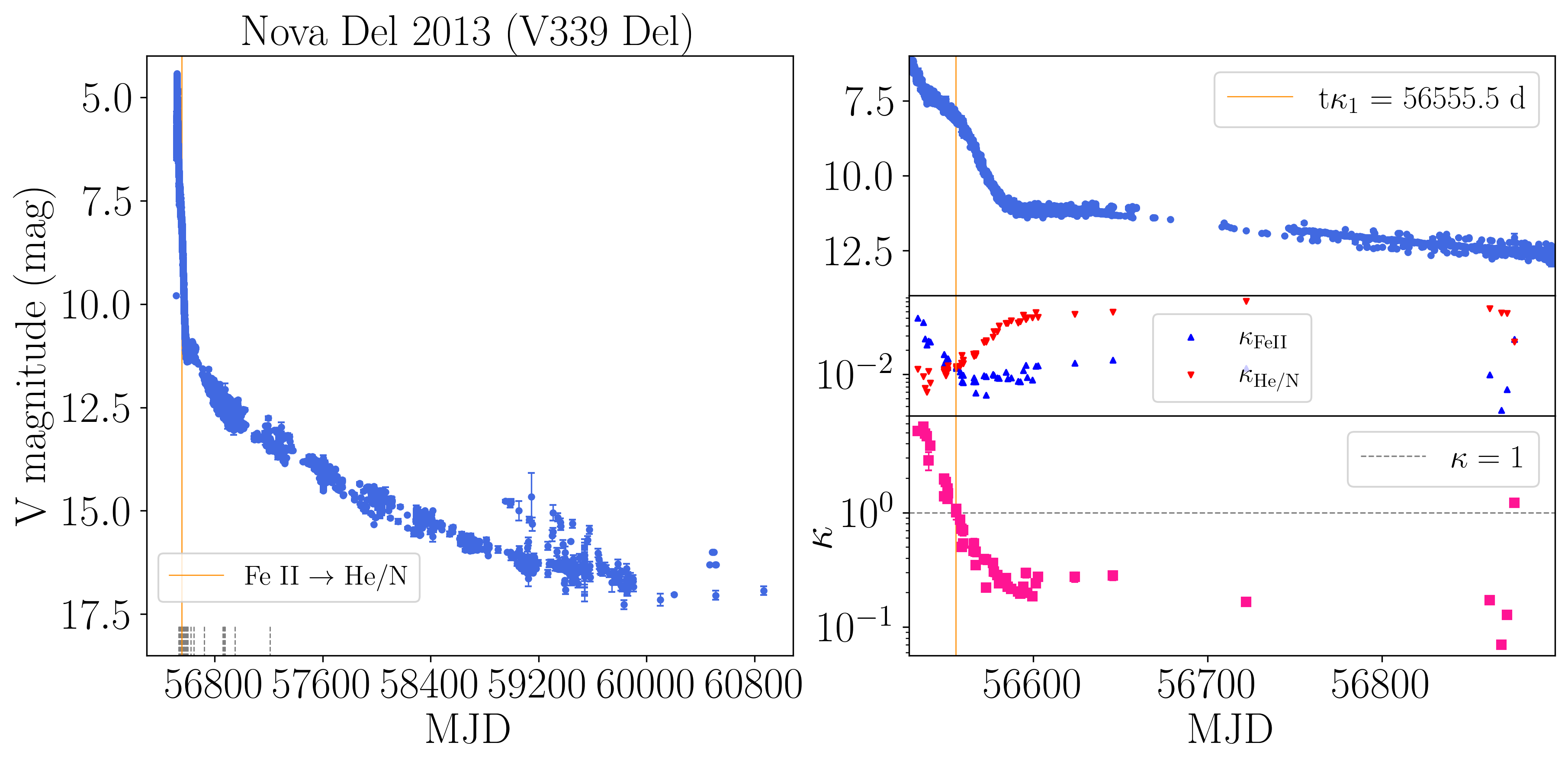}
    \includegraphics[width=\linewidth]{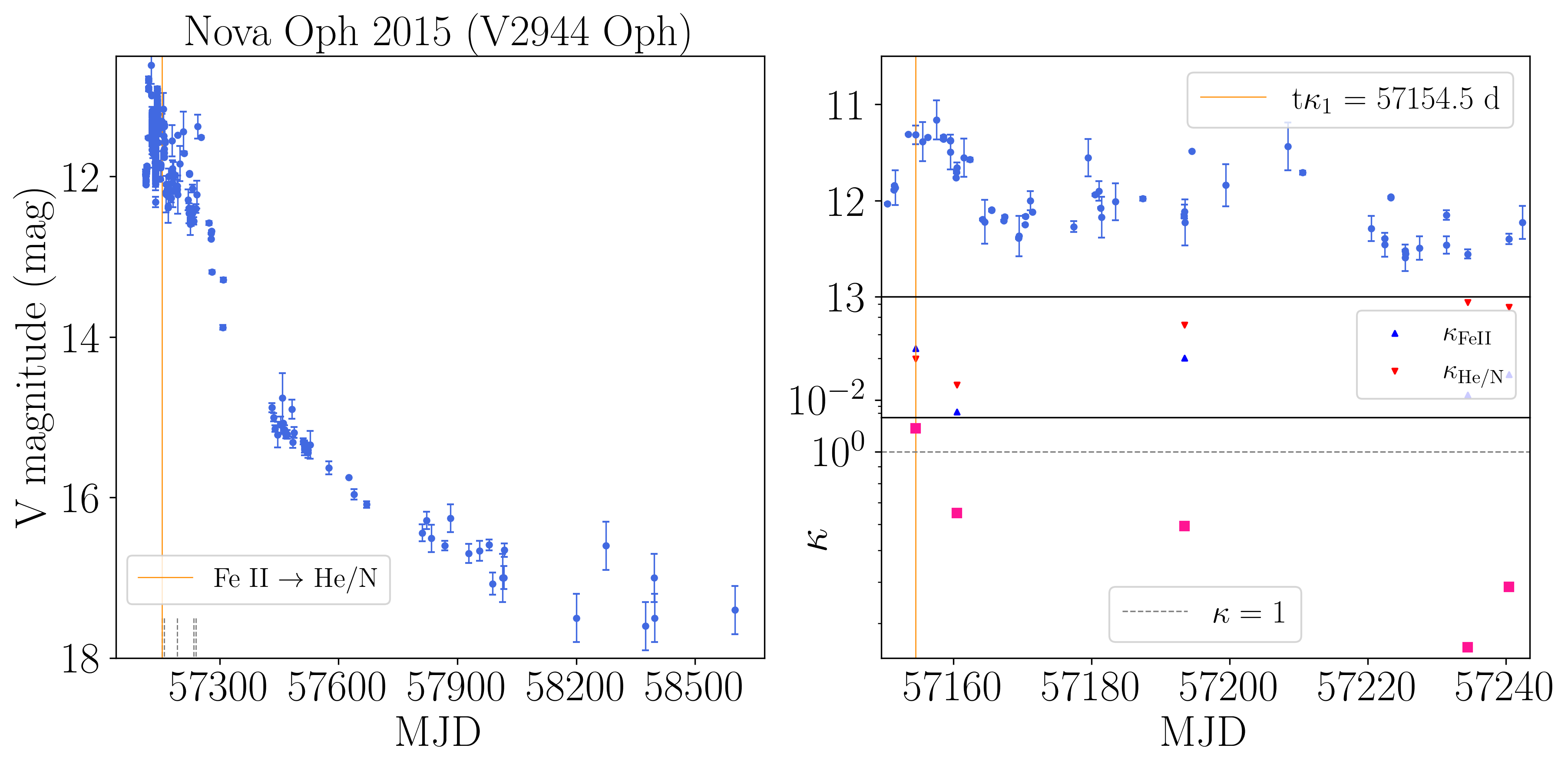}
    \includegraphics[width=\linewidth]{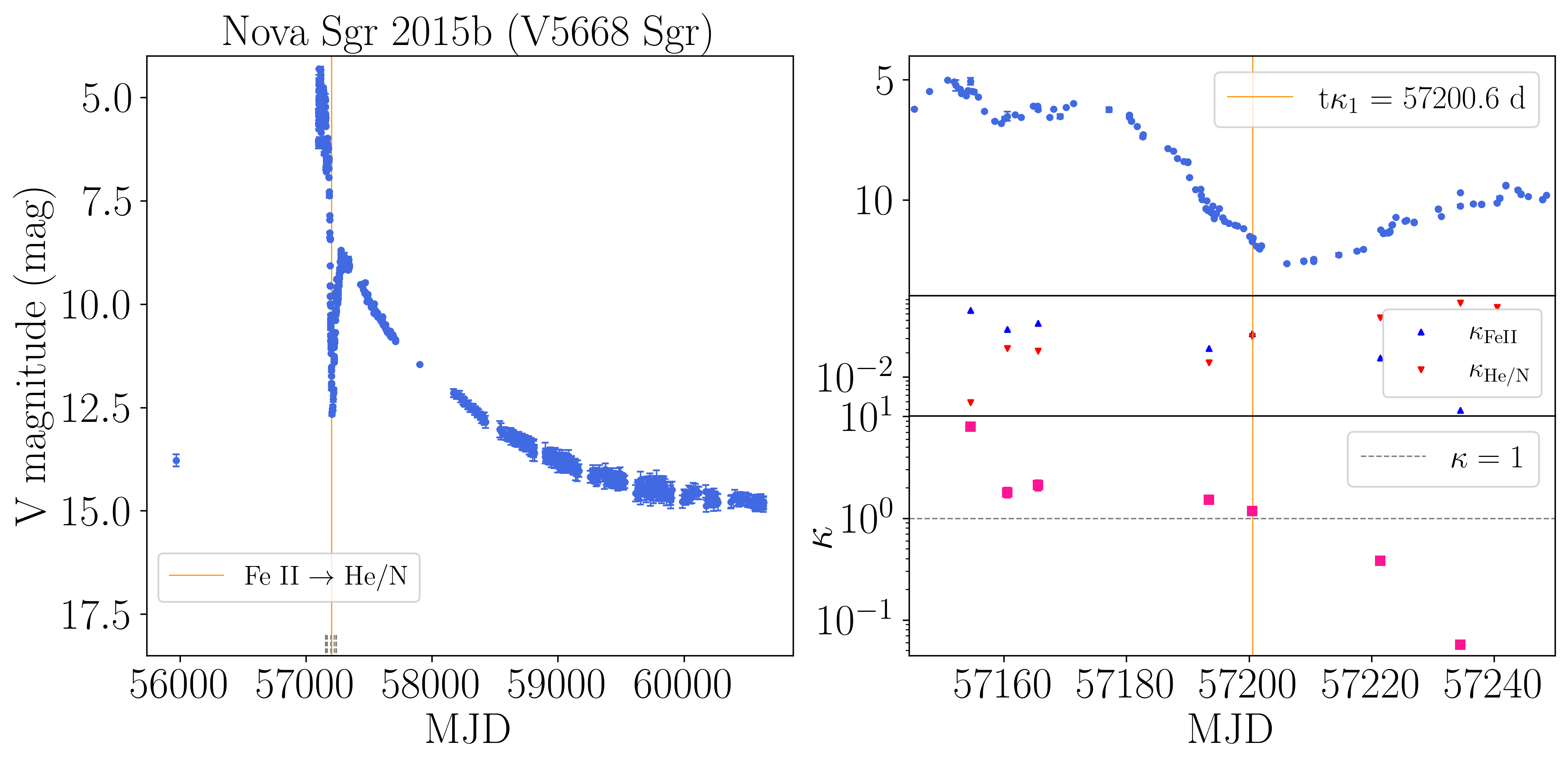}
    \includegraphics[width=\linewidth]{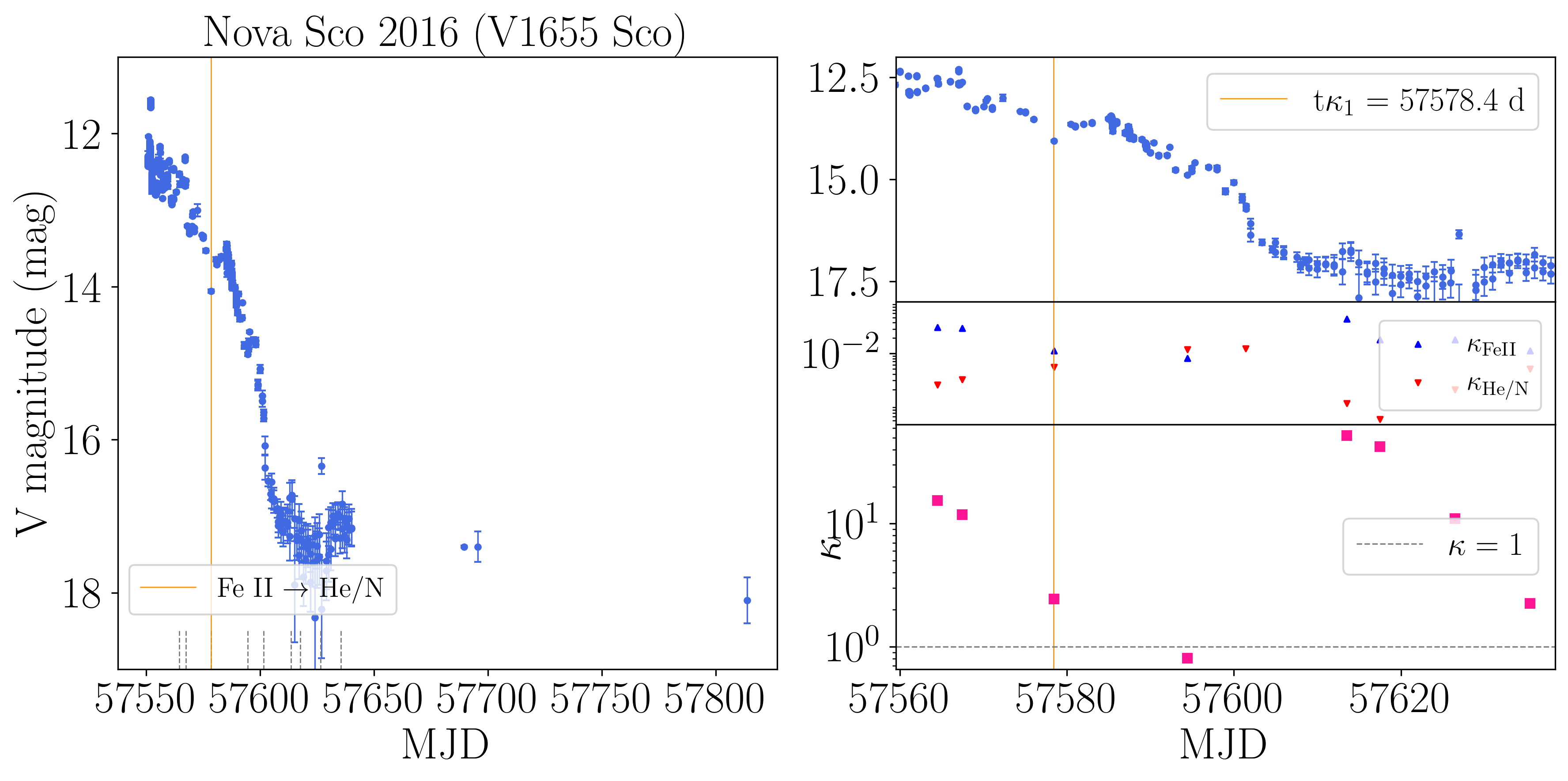}
    \includegraphics[width=\linewidth]{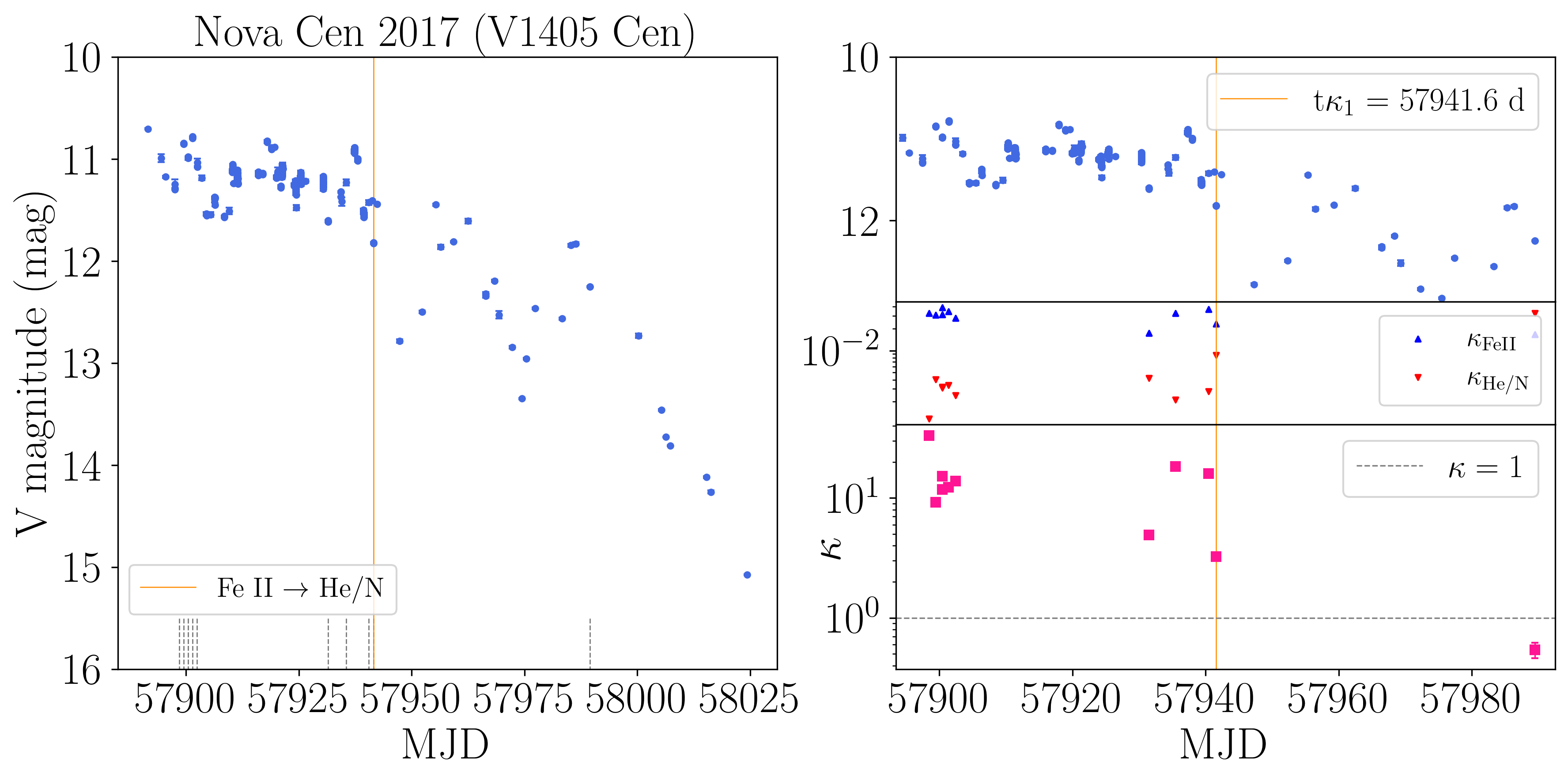}
\end{minipage}
\hfill
\begin{minipage}{0.46\textwidth}
    \centering
    \includegraphics[width=0.95\linewidth]{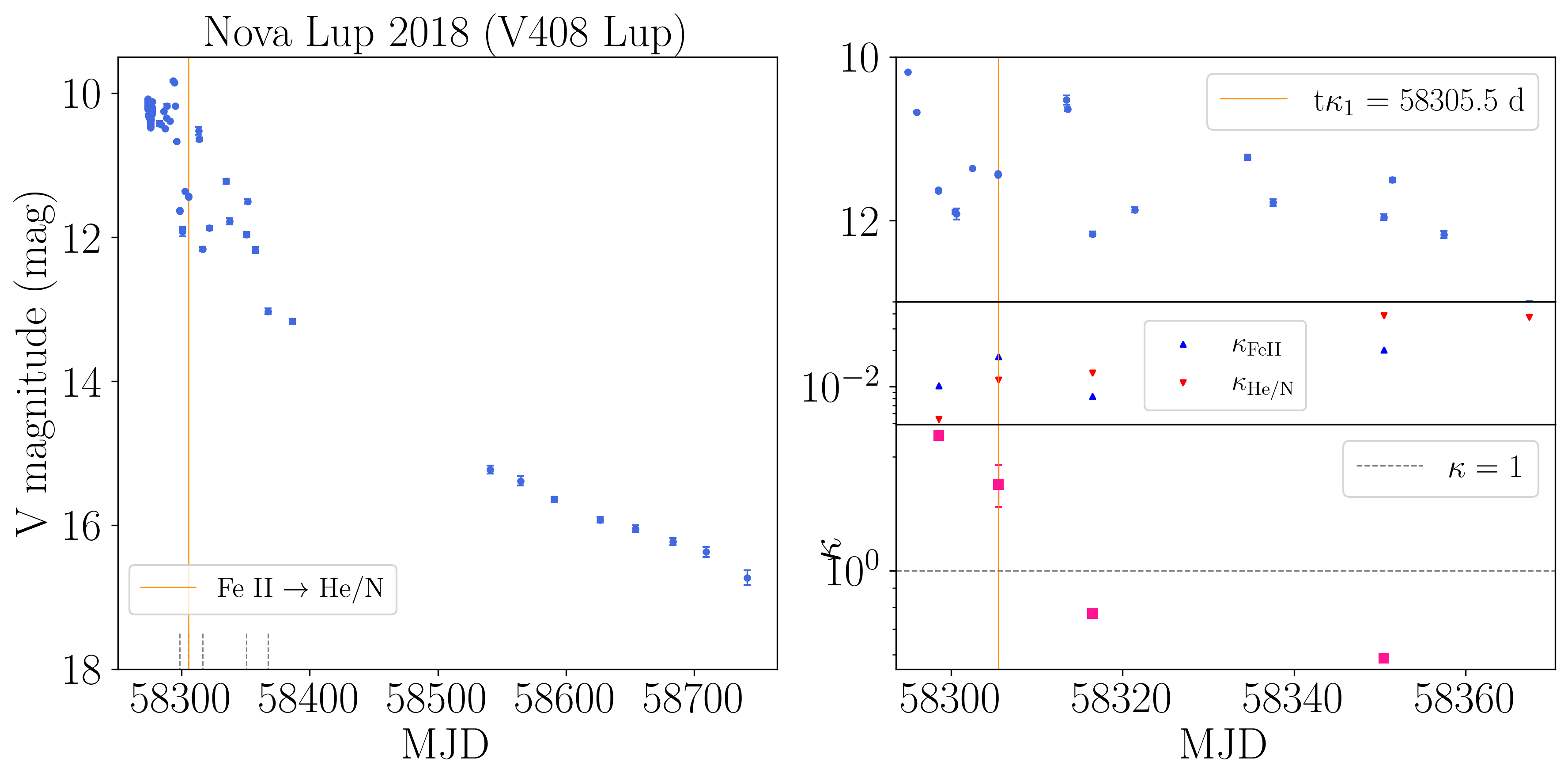}
    \includegraphics[width=0.95\linewidth]{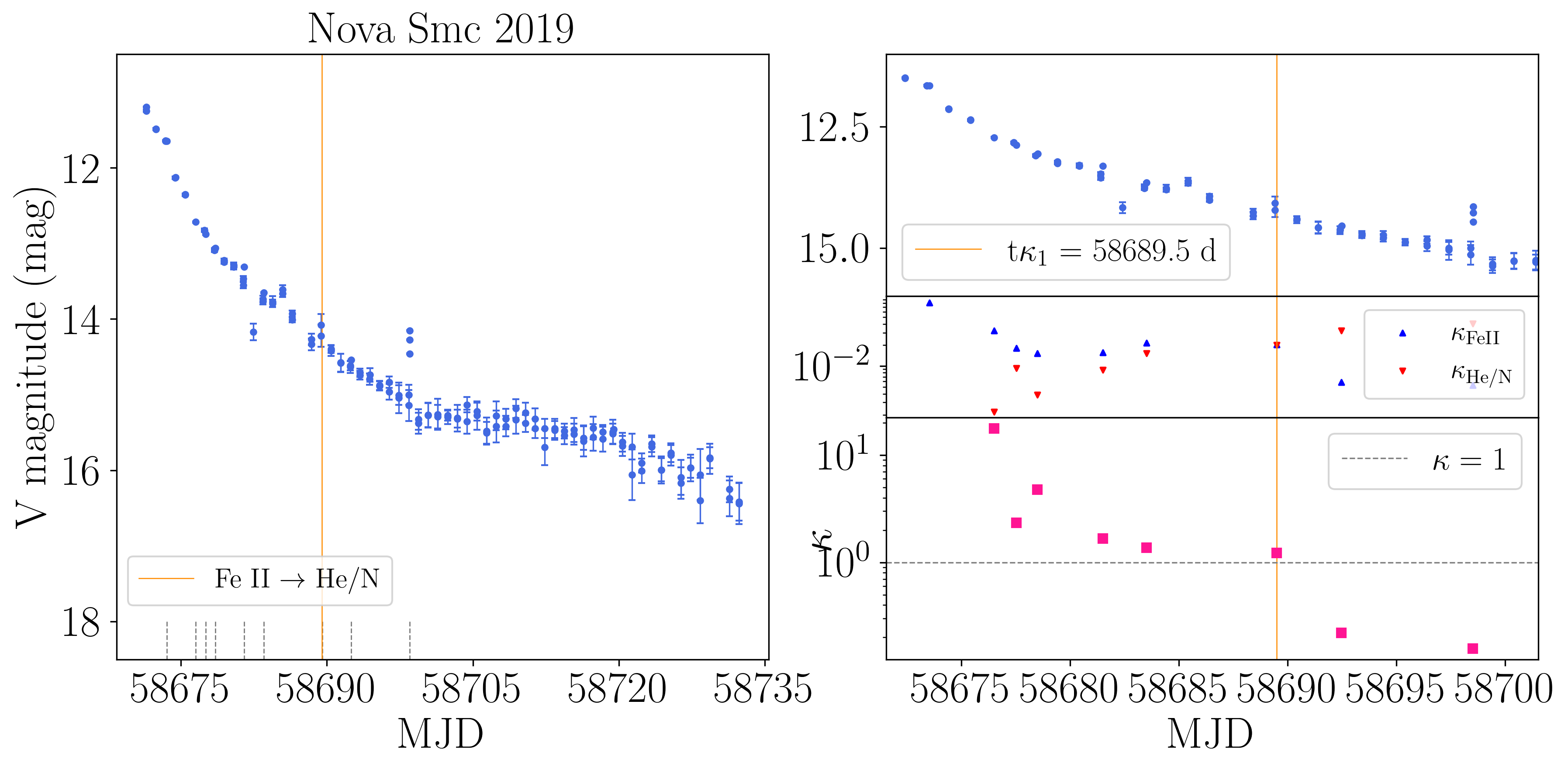}
    \includegraphics[width=0.95\linewidth]{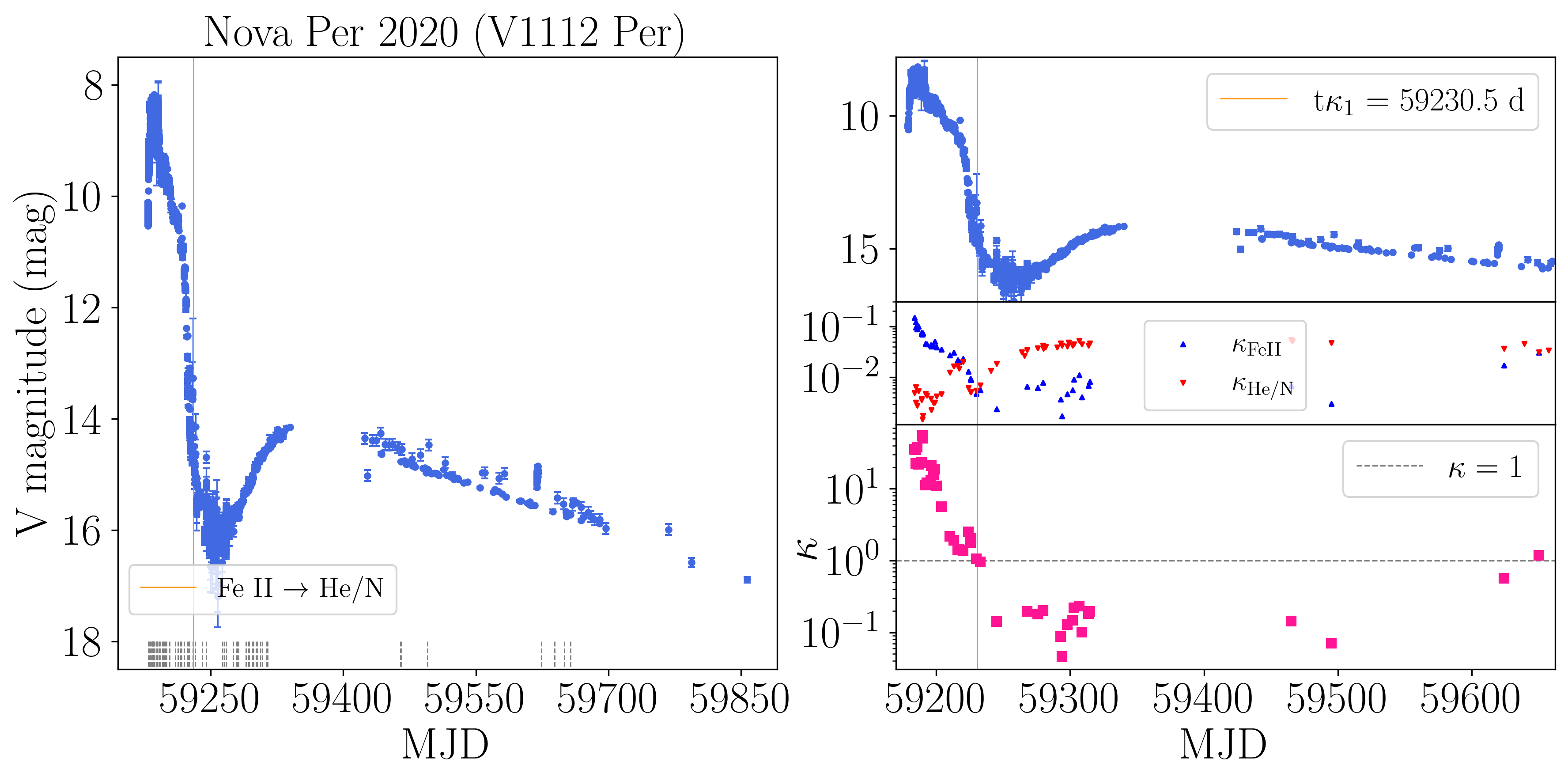}
    \includegraphics[width=0.95\linewidth]{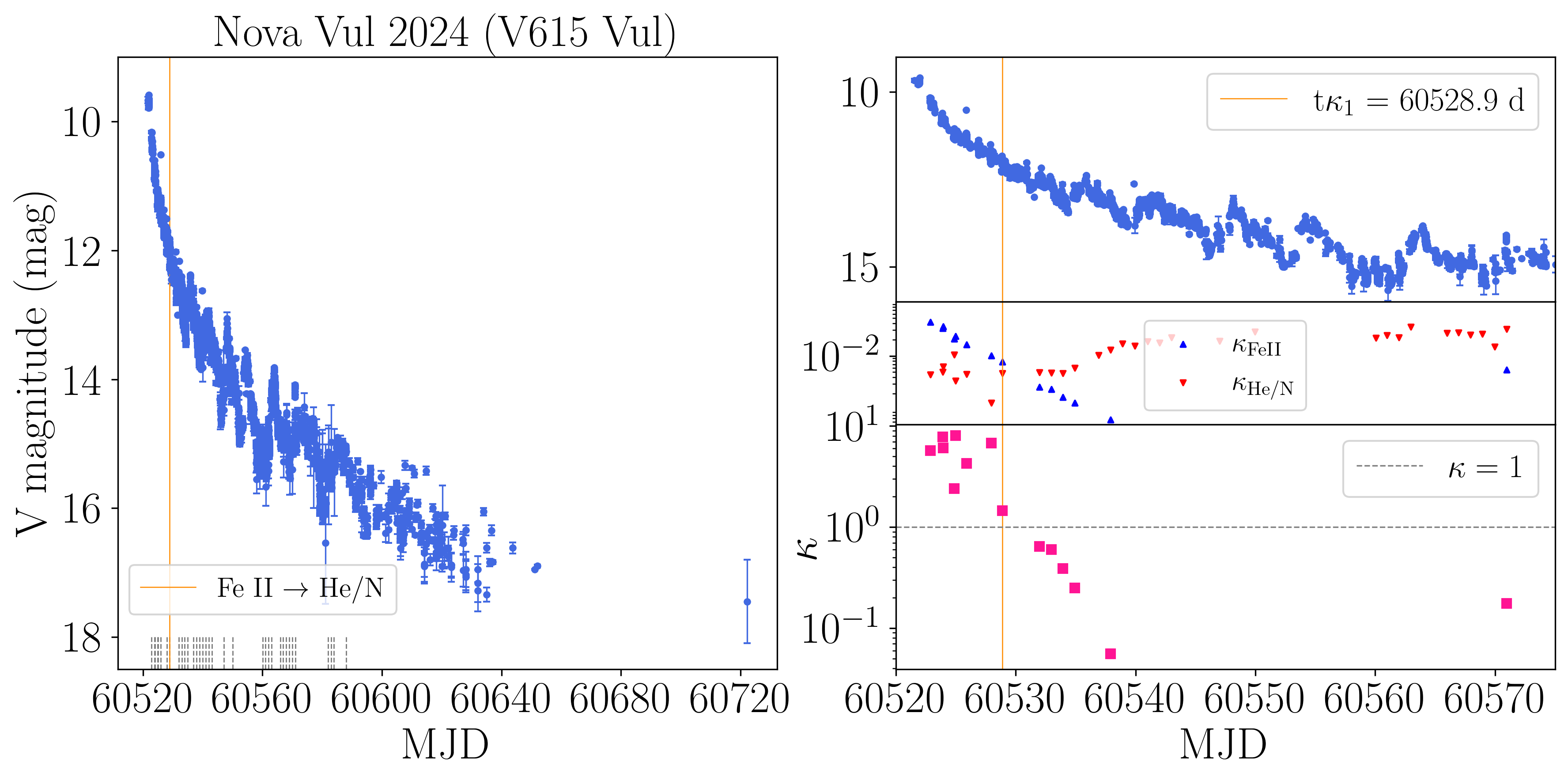}
    \includegraphics[width=0.95\linewidth]{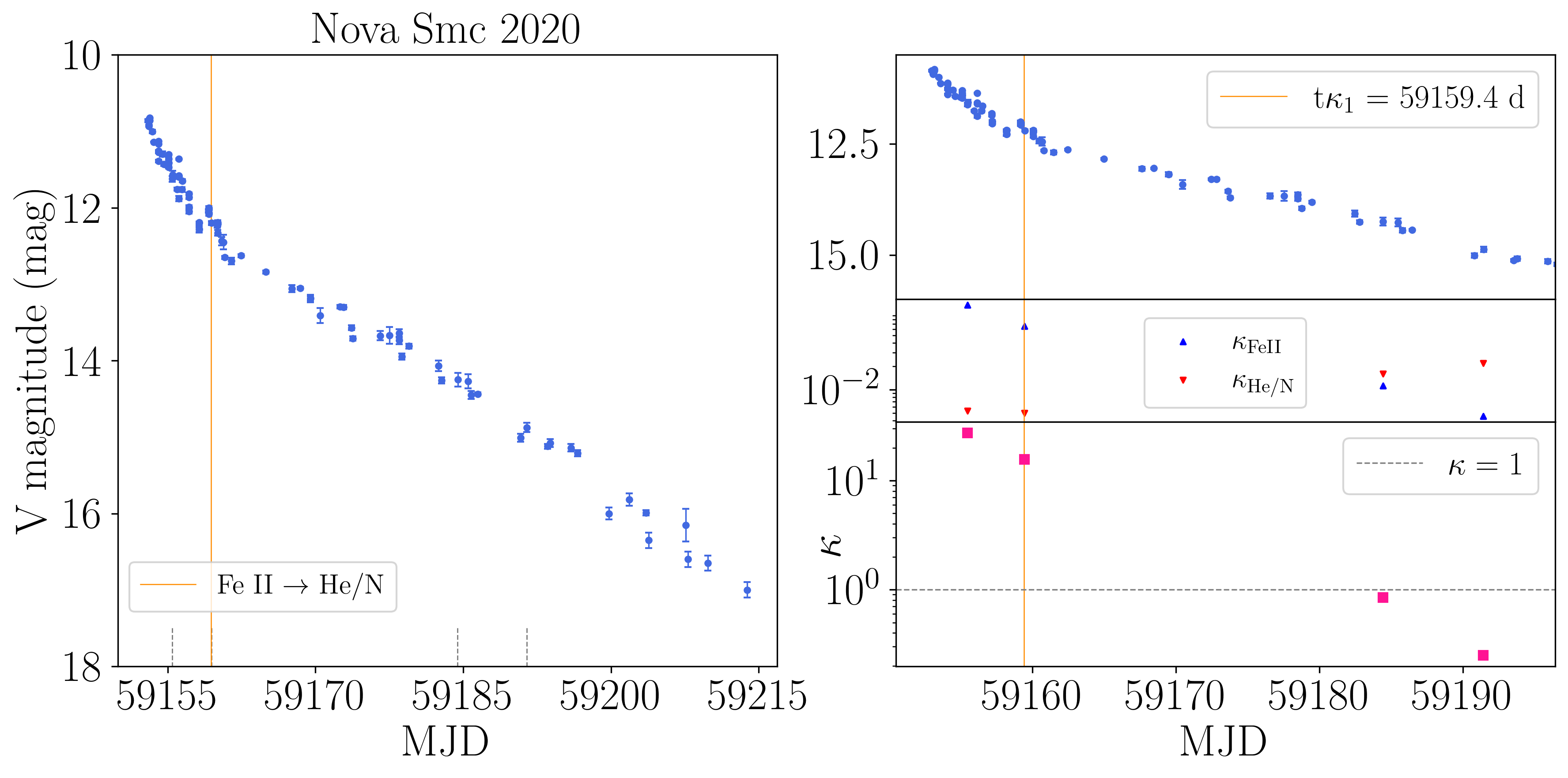}
\caption{
Photometric and spectral evolution of the novae that show a transition from \ion{Fe}{2} to He/N classes. 
For each nova, the left panel displays its AAVSO $V$ light-curve, and the right panels display zoomed-in views of the $V$ light-curve (\emph{upper}), of the $\kappa_{FeII}$ (blue) and $\kappa_{He/N}$ (red) coefficients (\emph{middle}), and the $\kappa$ coefficient (\emph{bottom}). 
The grey dashed lines indicate the epochs of the available spectra.
}
    \label{fig:allfigs}
\end{minipage}
\end{figure*}

For a significant number of novae, there is a noticeable time evolution of $\kappa$ from an \ion{Fe}{2}-dominated phase with $\kappa > 1$ to a He/N-dominated phase with $\kappa < 1$ (see Fig.~\ref{fig:allfigs}). 
Accordingly the $\kappa_\mathrm{Fe}$ (blue) and $\kappa_\mathrm{He/N}$ (red) coefficients exhibit a characteristic “scissors” pattern, reflecting the transition from an \ion{Fe}{2}-dominated to a He/N-dominated phase, as $\kappa$ declines below unity.

Let's consider the case of Nova Cen 2013 (top-left panels of Fig.~\ref{fig:allfigs}) to illustrate this behavior. 
It is a moderately fast nova with $t_2 = 37$ d and $t_3 = 65$ d with a D-shaped light-curve.  
A careful inspection of its light-curve reveals that the early evolution presented a short-lived plateau-like phase, that the dust dip is quite subtle, and that the late fading is pretty smooth and featureless.  
The first spectrum, obtained about 20 d before $t_2$, is \ion{Fe}{2}-dominated ($\kappa > 1$).  
Since then, the value of $\kappa$ declined monotonically to become $< 1$ (i.e.\ He/N-dominated) about 4 d after $t_3$ (marked by a vertical orange line in Fig.~\ref{fig:allfigs}), reaching its minimum a bit later, about 41 d after $t_3$. 
Since then, the values of $\kappa$ are consistently lower then unity.

This behavior of $\kappa$ is similar to that of novae in Fig.~\ref{fig:allfigs}, particularly for novae with adequate time coverage of the spectral data: Del\,2013\footnote{The last available spectrum of nova Del\,2013 seems to imply that $\kappa$ has recovered to a value close to unity, but it might be caused by its low S/N ratio.}, Sgr\,2015b, Sco\,2016, Lup\,2018, SMC\,2019, SMC\,2020, Per\,2020, and Vul\,2024. 
In many cases, these novae show D-class light-curves as Nova Cen\,2013, with a sharp early decline and a subsequent more gradual long-term brightness decline, although some of them exhibit jitters or oscillations in their light-curves. 
The time evolution of $\kappa$ mimics to some degree the behavior of the light-curves of these novae. 
It is above unity when the novae are bright, and below unity when the ligh-curve fades.

\subsubsection{Multiple transitions from \ion{Fe}{2} to He/N: \ion{Fe}{2} $\leftrightarrows$ He/N. }

\begin{figure*}
    \centering
 \includegraphics[width=0.65\linewidth]{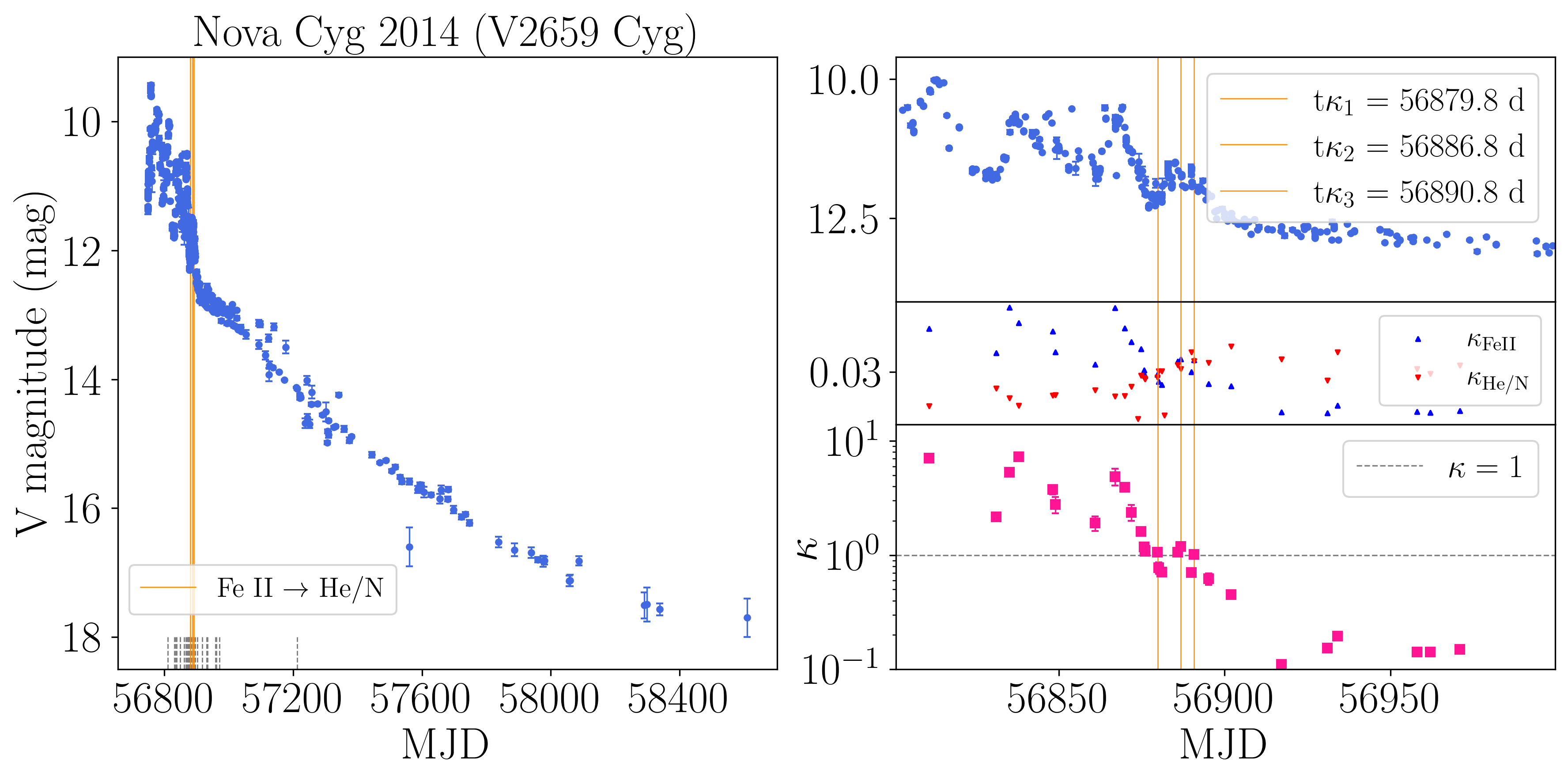}
 \includegraphics[width=0.65\linewidth]{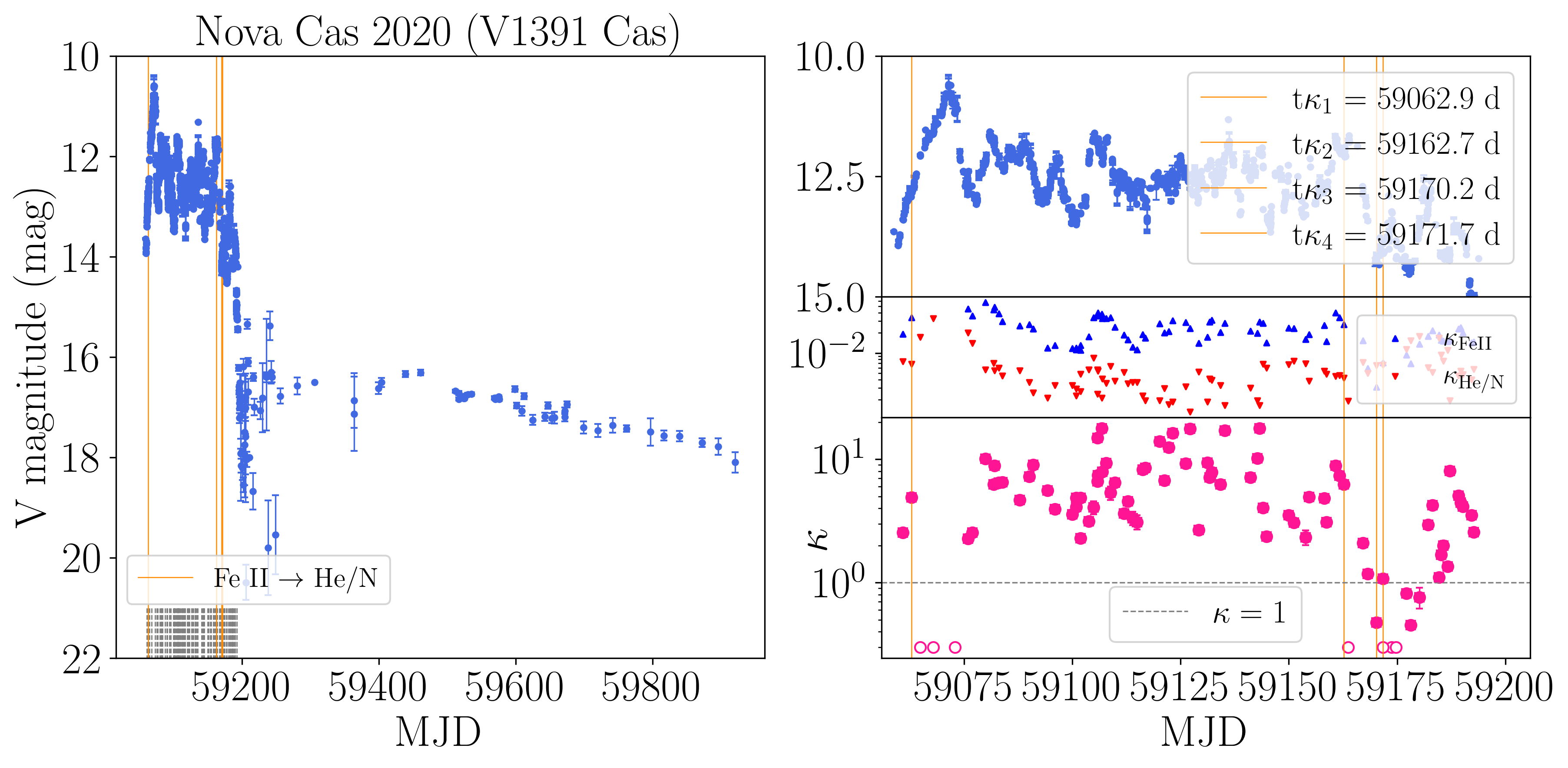}
 \includegraphics[width=0.65\linewidth]{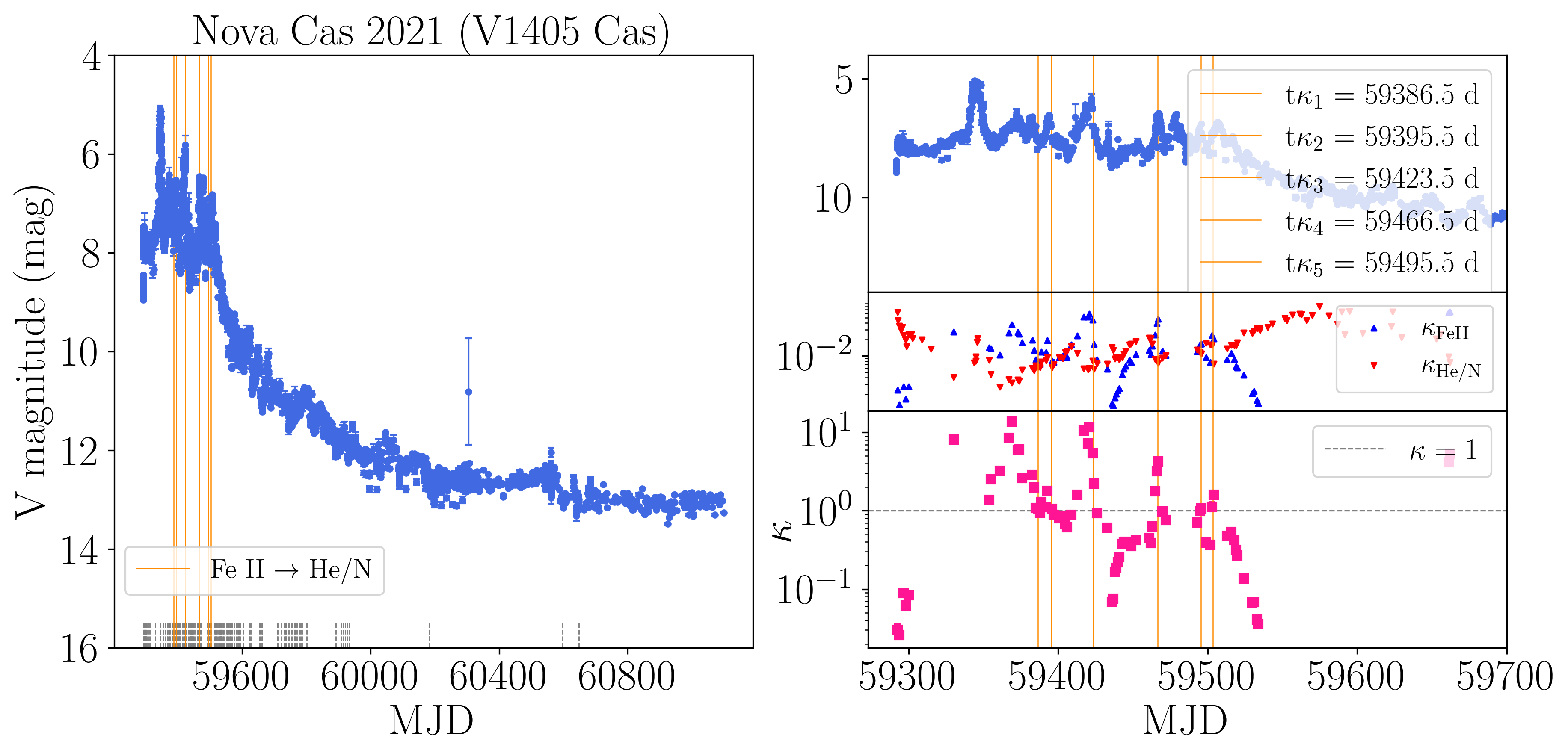}
 \includegraphics[width=0.65\linewidth]{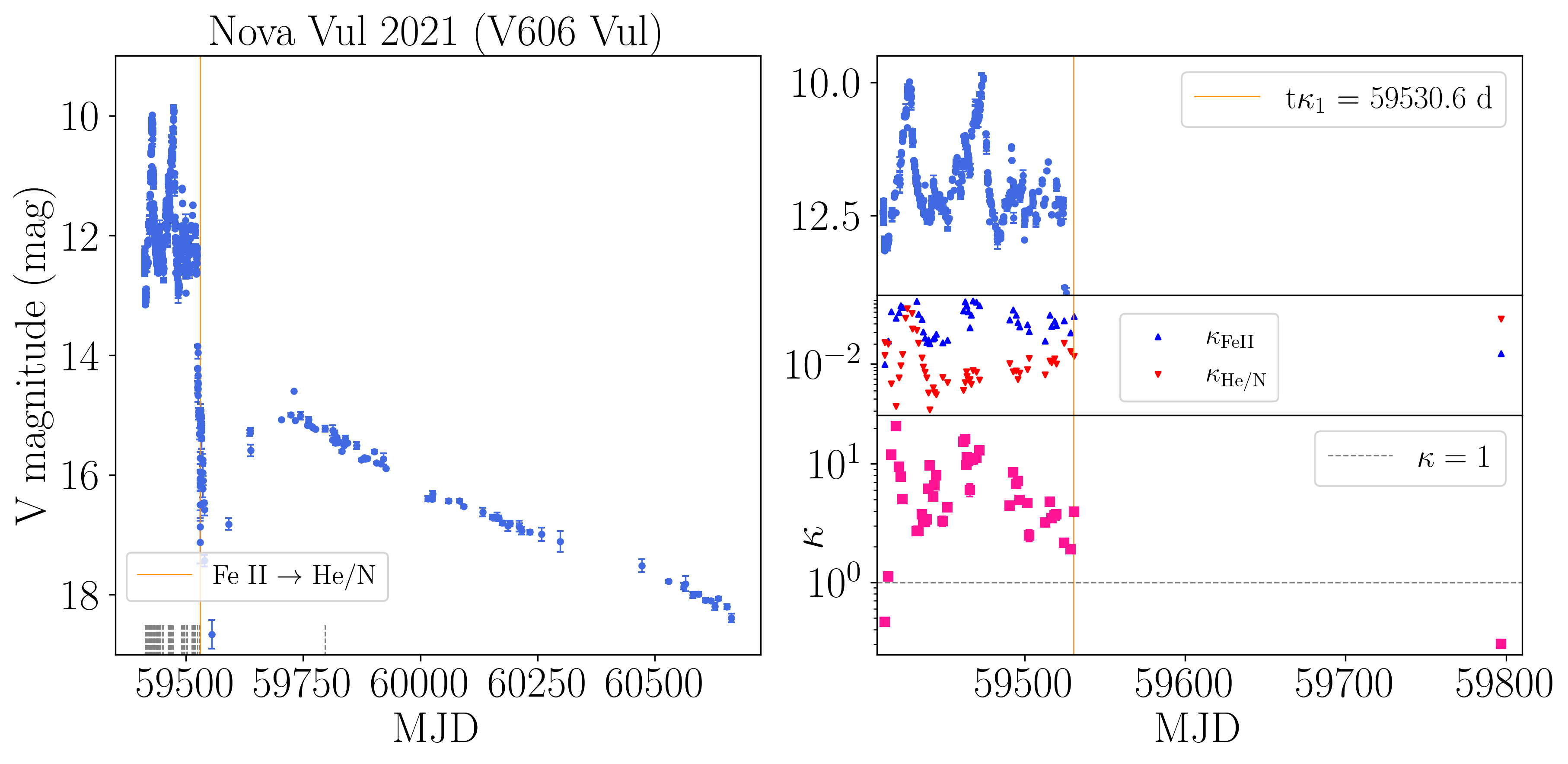}
\caption{
Same as Fig.~\ref{fig:allfigs} but for the novae of our main sample that present oscillatory behavior (jitters). 
The open circles in the lower right panel of Nova Cas\,2020 correspond to spectra with null flux of the \ion{Fe}{2} line. 
Since these values do not appear in logarithmic scale, we replaced them with very low values in order to show the transition from He/N to \ion{Fe}{2} phase and vice versa.
}
\label{fig:jitters}
\end{figure*}

For a few novae, the time evolution of $\kappa$ shows multiple transitions from values higher than unity to values below it and back  (Fig.~\ref{fig:jitters}). 
These transitions, which occur on short time-scales and might be multiple, are also linked to the light-curve.

Let's consider the case of Nova Cas\,2021 to illustrate this category. 
This is a very slow nova ($t_2 = 166$ d, $t_3 = 175$ d) with a J-class light-curve. 
Actually the light-curve shows a relatively flat plateau phase superposed by erratic jitters and a subsequent long-term decline. 
In this nova, $\kappa$ shows an increase from a value in the first available spectrum $< 1$, to then fluctuate up and down, marking successive \ion{Fe}{2}- and He/N-dominated phases. 
This behavior is caused mostly by large flux variations of the \ion{Fe}{2} emission line ($\kappa_\mathrm{Fe}$), whereas the He and N average flux value ($\kappa_\mathrm{He/N}$) remains much stable, with an initial decline trend followed by a more paused, but consistent increase. 
In this case again we observe that $\kappa$ closely follows the behavior of the light curve, with sharp increases of its value associated with specific jitters of its light-curve.  
As for the novae in the previous group, the value of $\kappa$ becomes smaller than unity once the nova enters the phase of gradual fading. 
This same behavior is shown by Nova Cyg\,2014,  Cas\,2020 and Vul\,2021. 
Despite the light-curve classification, all of them show plateau-like phases superposed by large amplitude jitters.

 \begin{figure*}
 \centering
 \begin{minipage}{0.48\linewidth}
     \centering
     \includegraphics[width=0.85\linewidth]{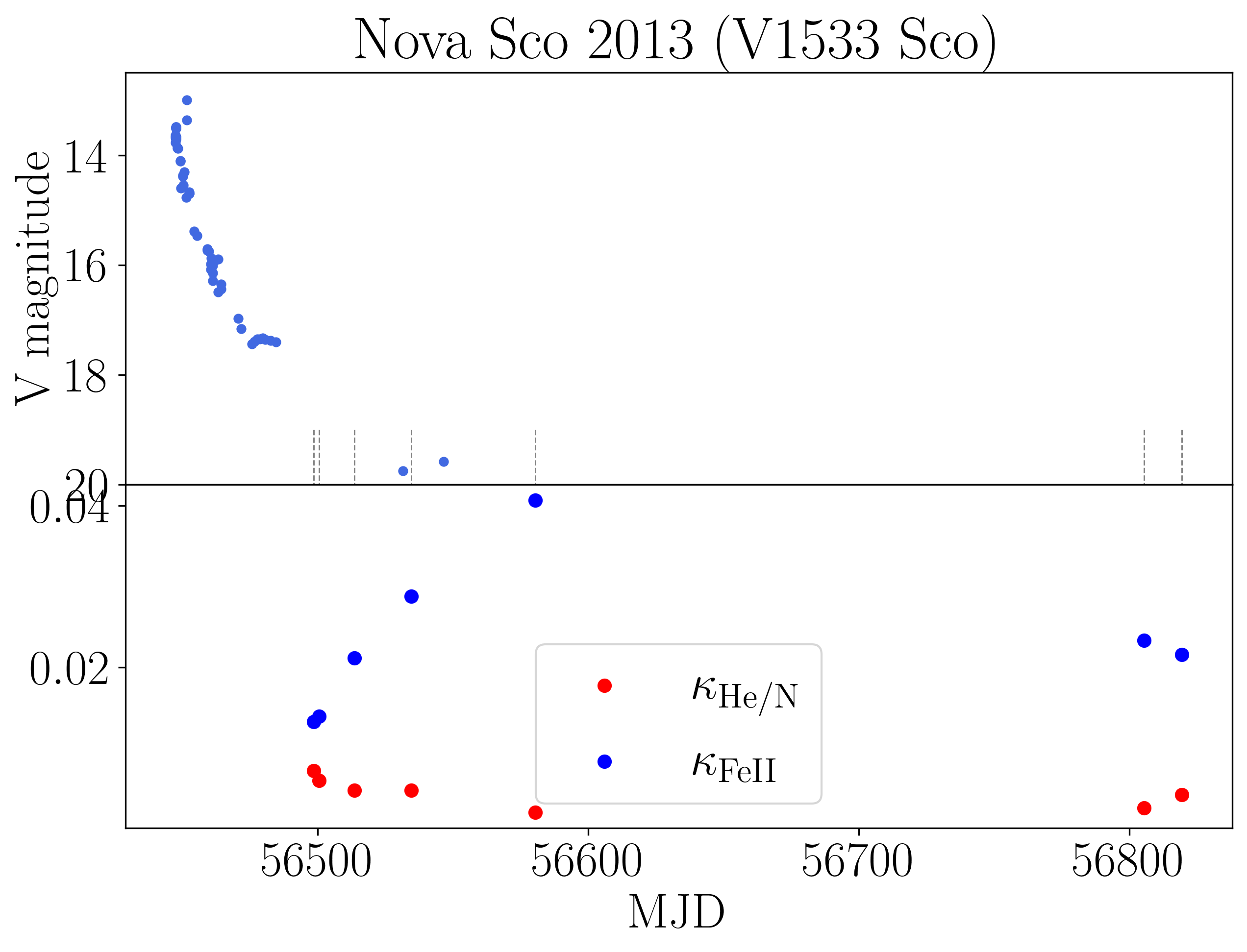}
     \includegraphics[width=0.85\linewidth]{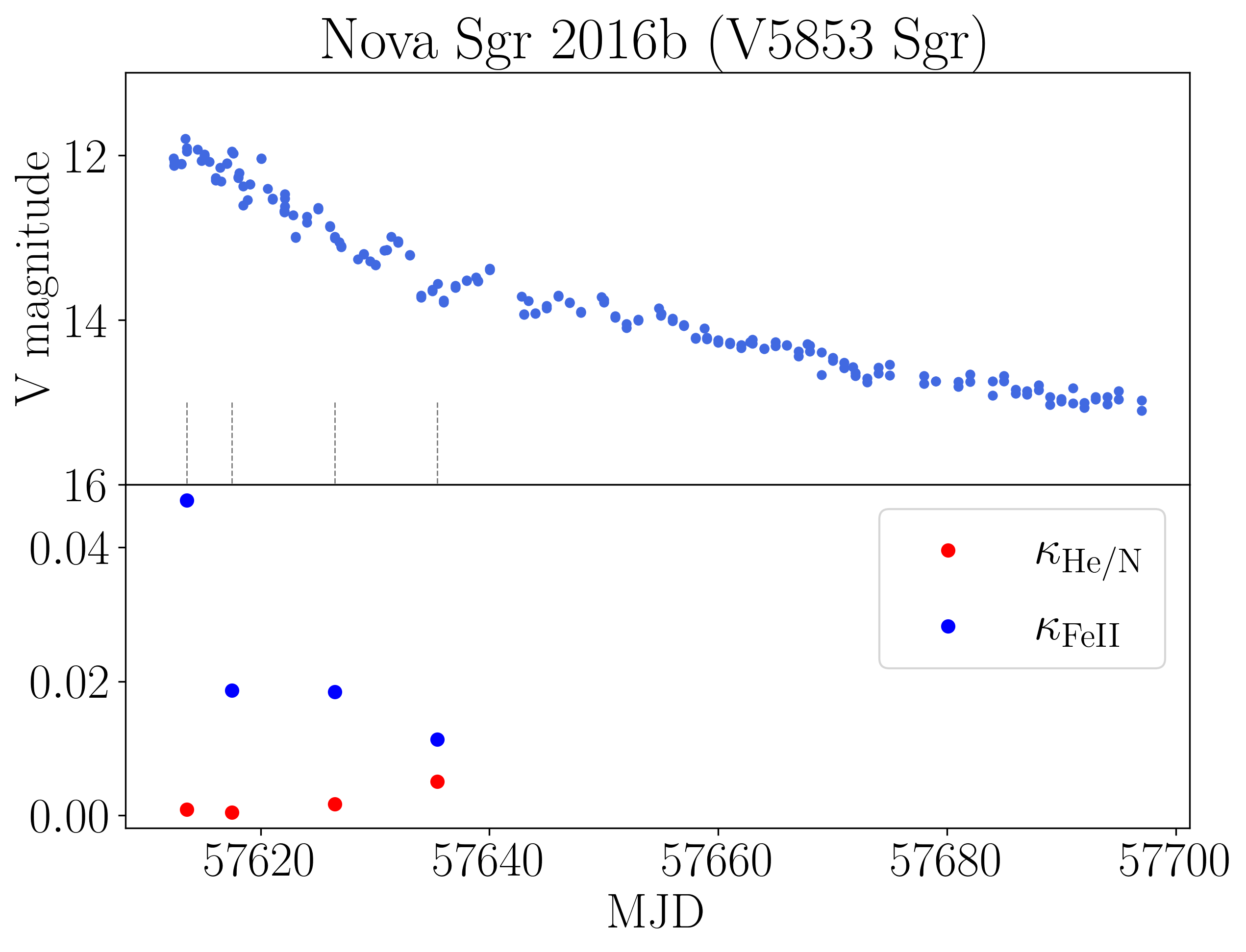}
    \includegraphics[width=0.85\linewidth]{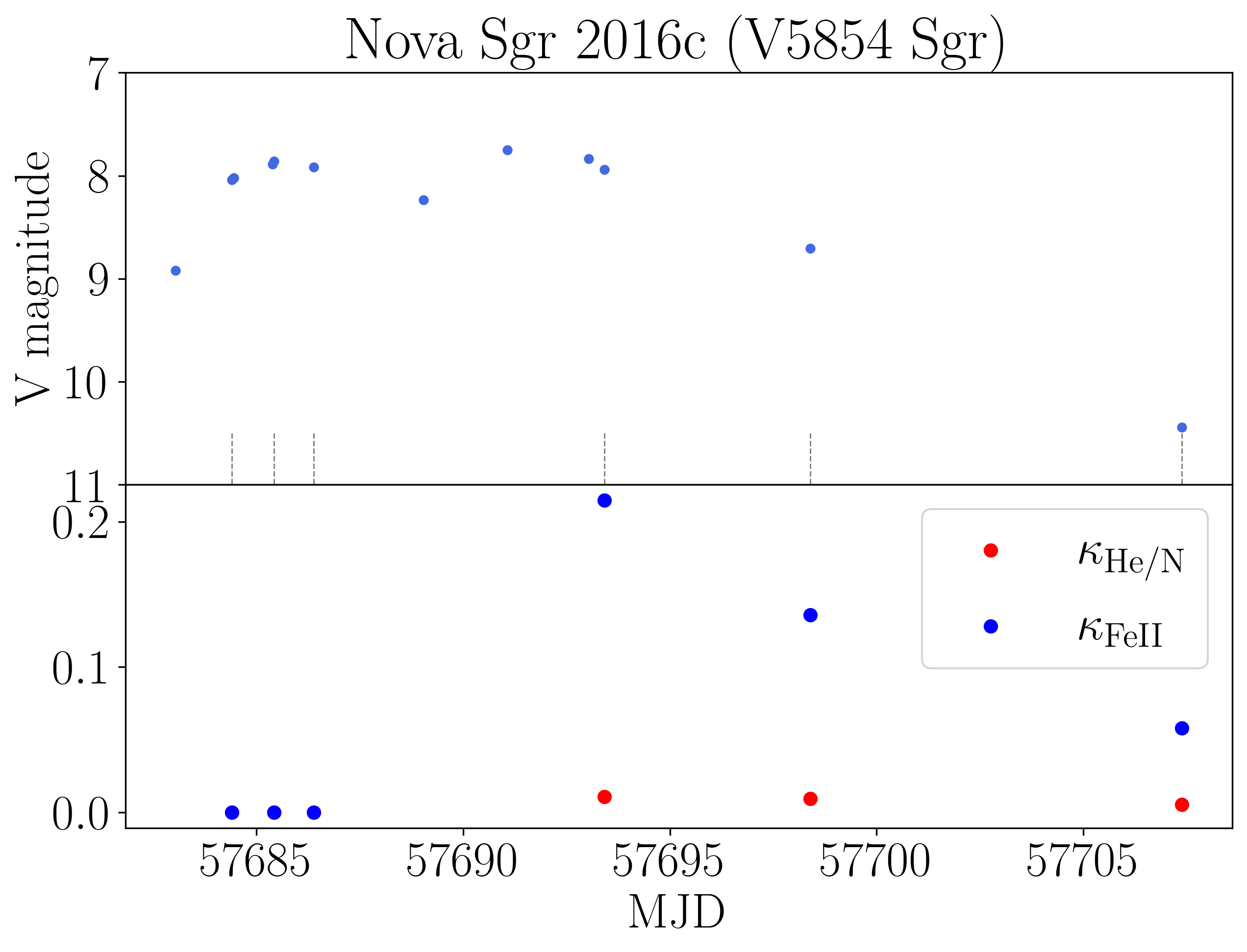}
    \end{minipage}
     \hfill
 \begin{minipage}{0.48\linewidth}
     \includegraphics[width=0.85\linewidth]{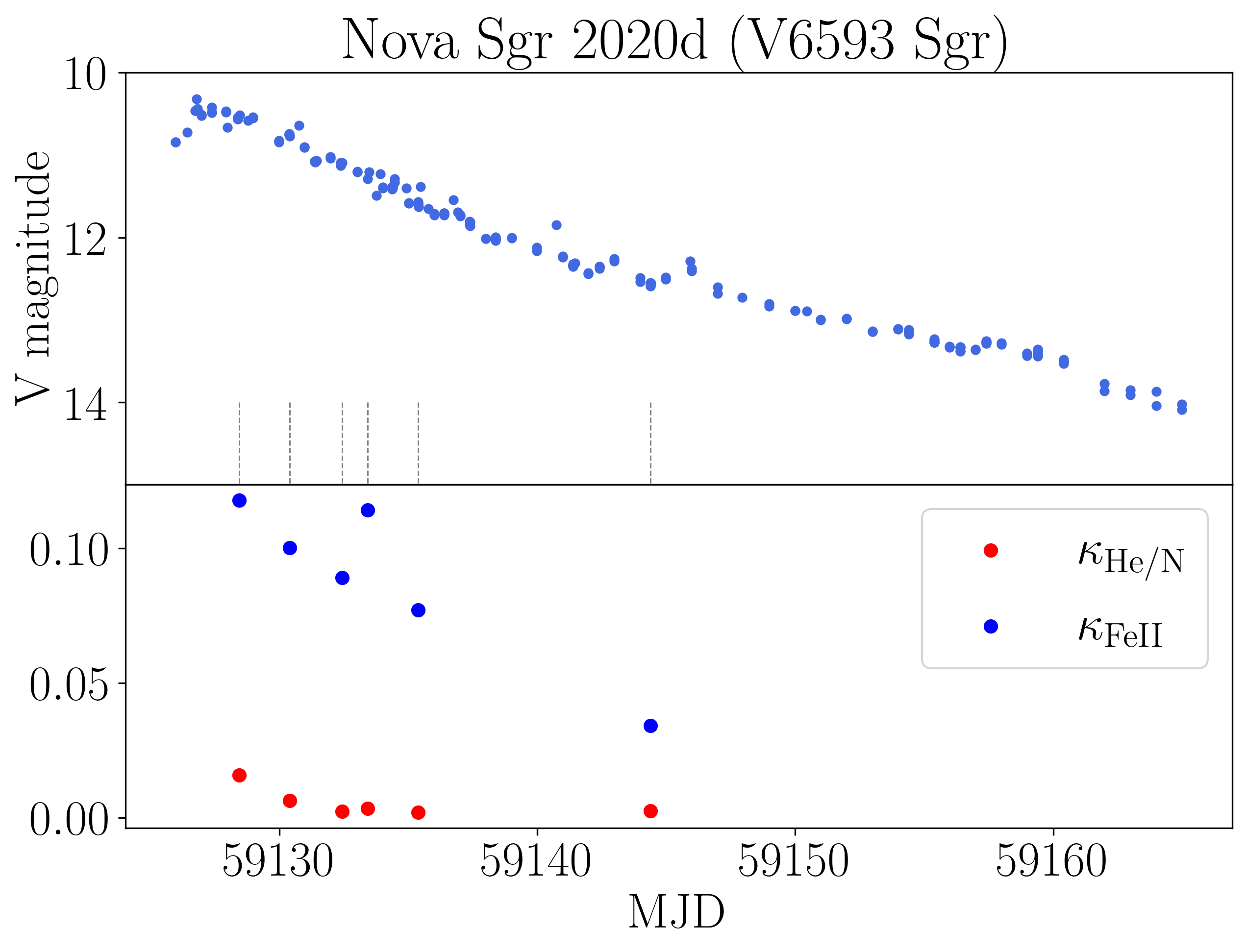}
     \includegraphics[width=0.85\linewidth]{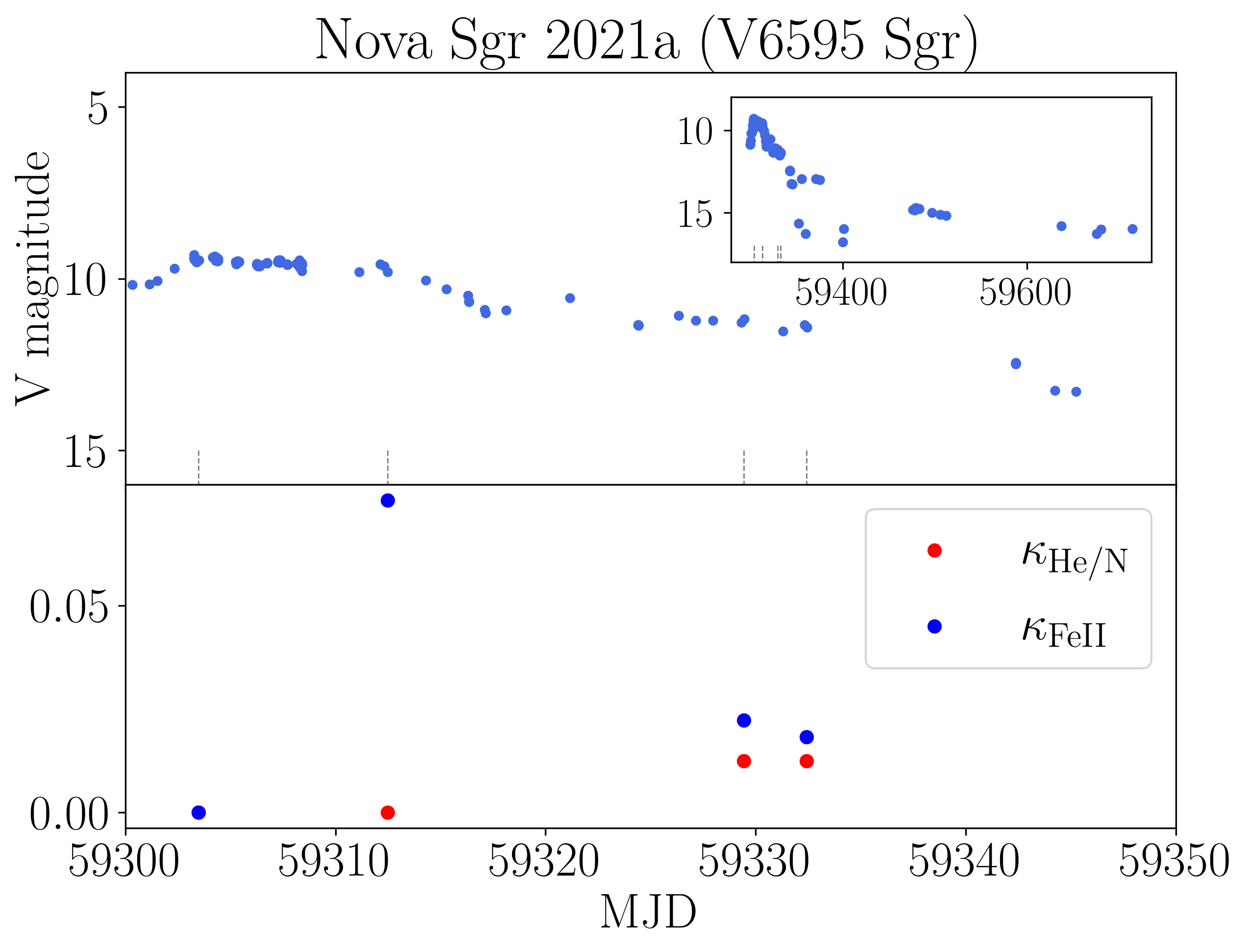}
     \includegraphics[width=0.85\linewidth]{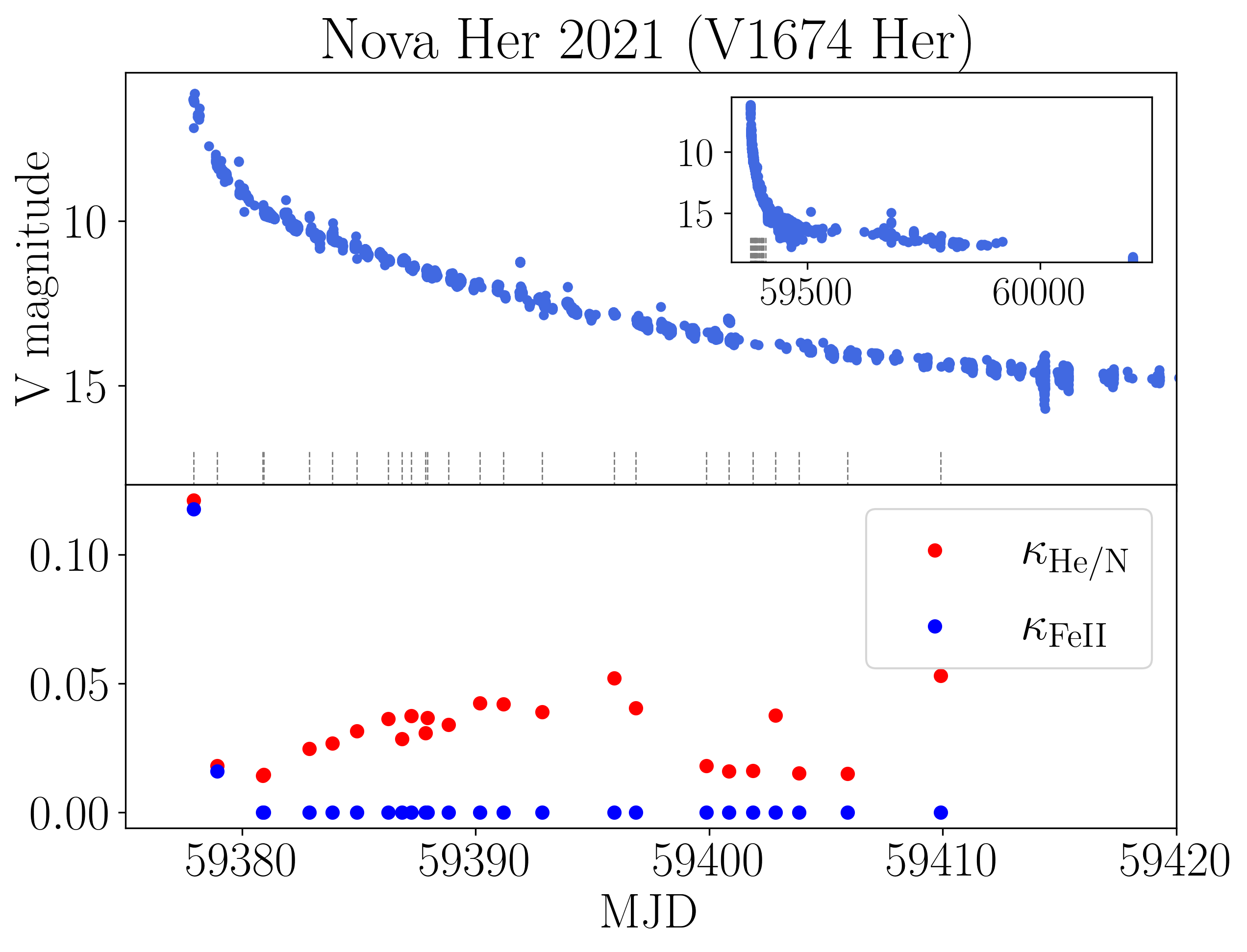}
 \end{minipage}
\caption{
{\it Upper panel:} 
$V$ light-curves of the novae that show a tendency to evolve from the \ion{Fe}{2}-dominated to the He/N-dominated phase, although $\kappa$ never crosses unity. 
{\it Lower panel:} Evolution of  $\kappa_{FeII}$ (blue) and $\kappa_{He/N}$ (red). 
The grey dashed lines indicate the epochs of the available spectra.
}
 \label{fig:tendency}
 \end{figure*}

\subsection{Transition tendency from \ion{Fe}{2} to He/N spectra and vice versa.}

Six novae in our sample (second block of Tab.~\ref{tab:novae}) show trends toward a transition from one spectral class to the other, with the value of $\kappa$ approaching, but not crossing, the $\kappa=1$ line (Fig.~\ref{fig:tendency}). 
Five of these novae are classified as \ion{Fe}{2}, however, their spectral coverage is incomplete and does not extend to late stages of the nova fading, when novae typically transition to the He/N spectral class. 
In contrast, Nova Her 2021 is classified as He/N because $\kappa$ is always smaller than unity.  
The two earliest available spectra of this nova show values of $\kappa$ close to one, with $\kappa_\mathrm{Fe}$ and $\kappa_\mathrm{He/N}$ apparently immediately after the ``scissor pivot''.  
This is a very fast nova ($t_2 = 1$ d) where \ion{Fe}{2} lines might have been dominant only in the very few hours after the outburst. 
This epoch was not covered by the available spectra.

\subsection{No transition from \ion{Fe}{2} to He/N, or vice versa.}

\begin{figure}
\centering
    \centering
    \includegraphics[width=0.45\linewidth]{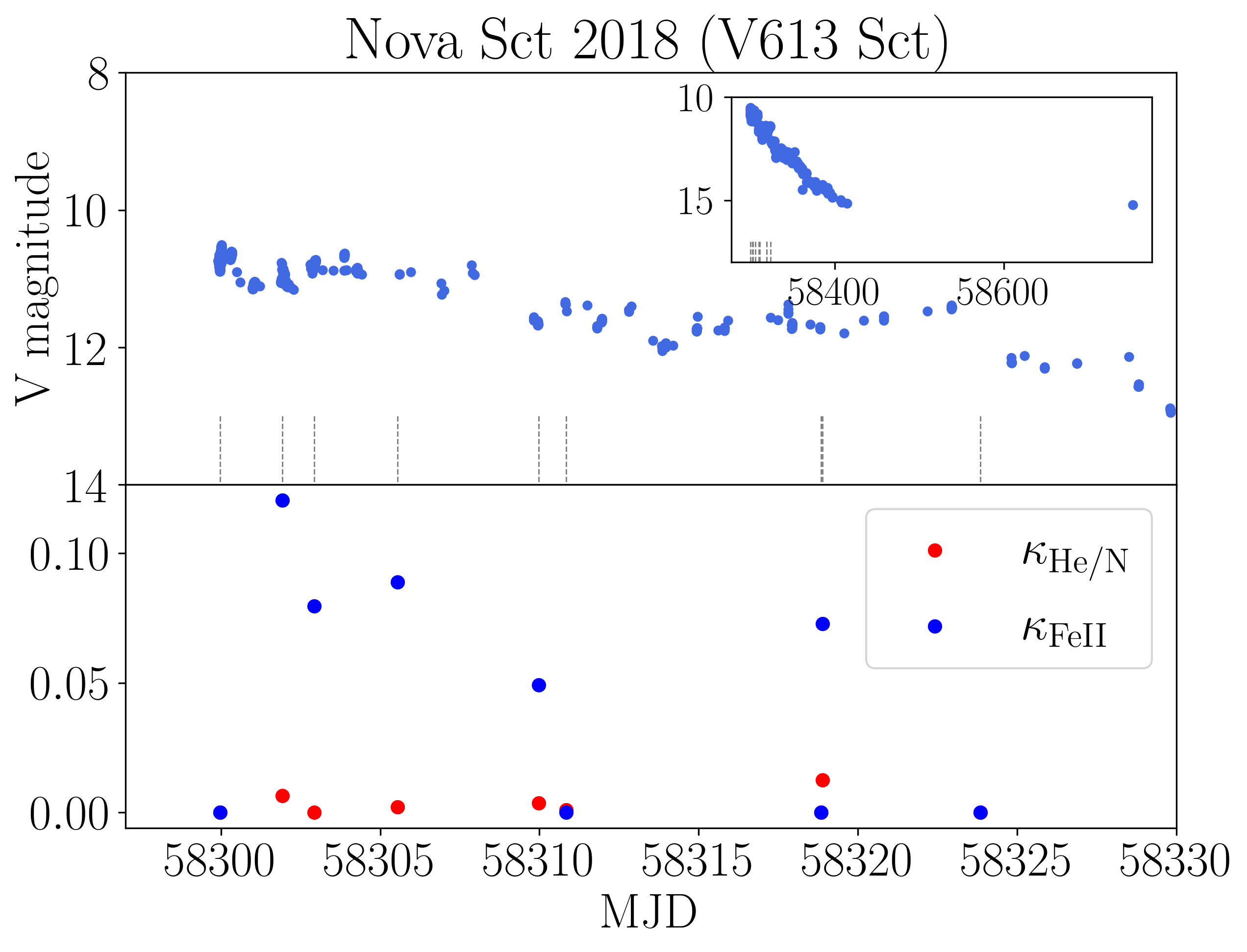}
    \includegraphics[width=0.45\linewidth]{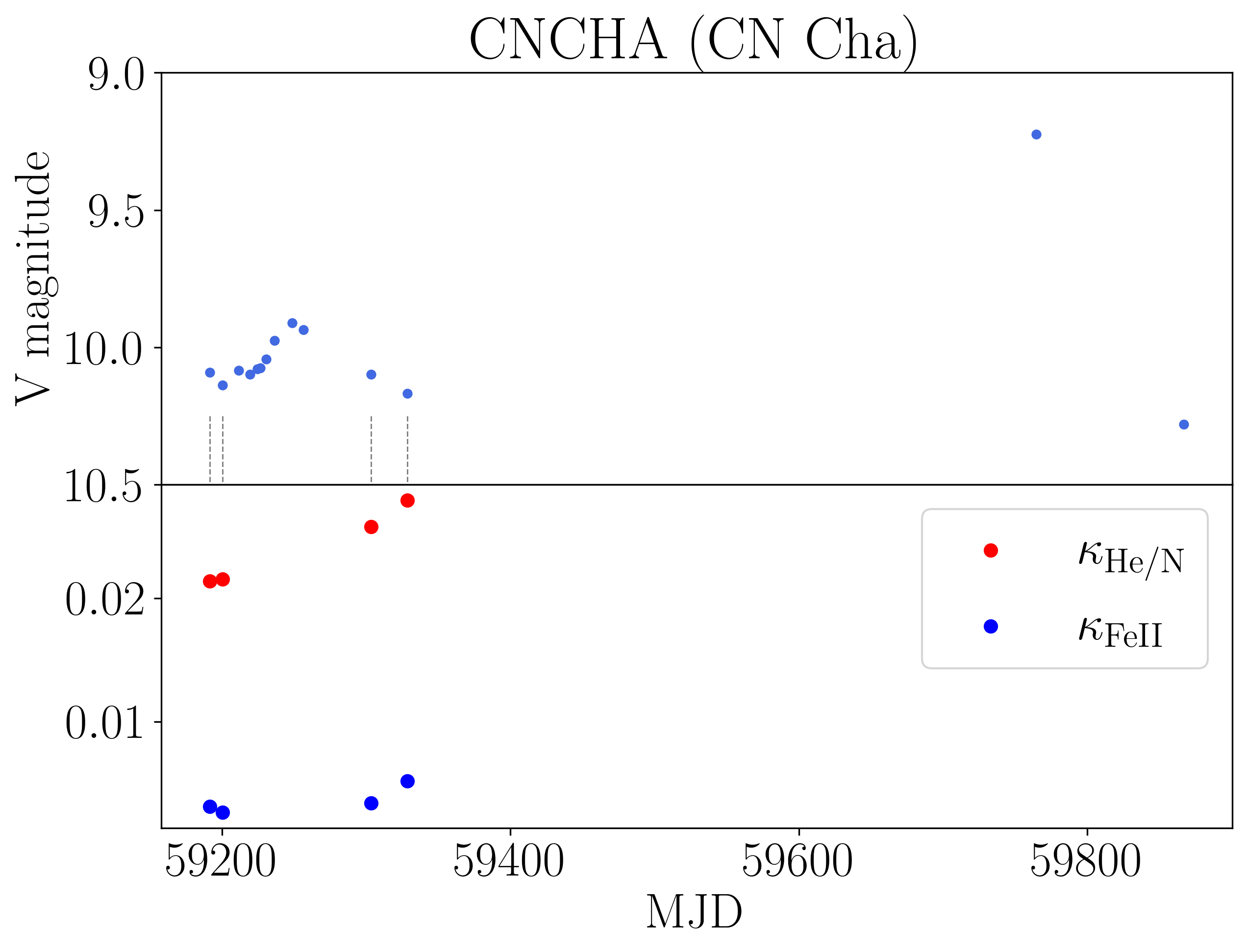}
    \includegraphics[width=0.45\linewidth]{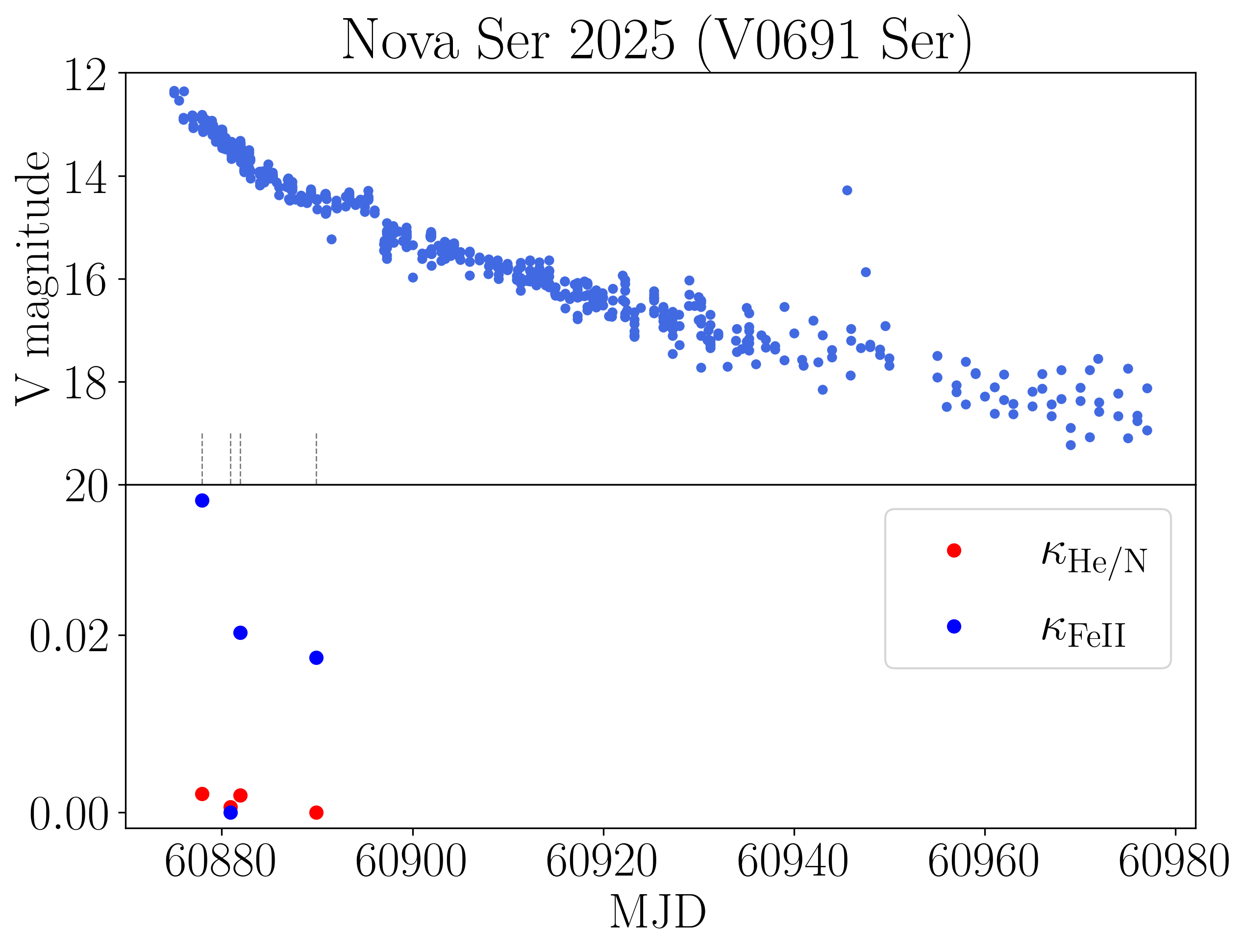}
\caption{
{\it Upper panel:} $V$ light-curves of the novae that do not show any tendency or evolution from the \ion{Fe}{2}-dominated to the He/N-dominated phase, probably due to limited spectral coverage. {\it Lower panel:} Evolution of  $\kappa_{FeII}$ (blue) and $\kappa_{He/N}$ (red). 
The grey dashed lines indicate the epochs of the available spectra.}
\label{fig:rest}
\end{figure}

There are three additional novae in our sample (third block of Tab.~\ref{tab:novae}) for which $\kappa$ does not show a clear trend (Fig.~\ref{fig:rest}).  
The time coverage of the spectra of novae Sct\,2018 and Ser\,2025 seems to be limited to the early phases, with both being in the \ion{Fe}{2} phase. 
Otherwise CN\,Cha is a very slow symbiotic nova \citep{2020AJ....160..125L} with poor time coverage.  
Apparently there was a brightness peak between the second and third available spectra that was therefore spectroscopically missed.  

\subsection{Recurrent novae}

\begin{figure}
\centering
    \centering
    \includegraphics[width=0.6\linewidth]{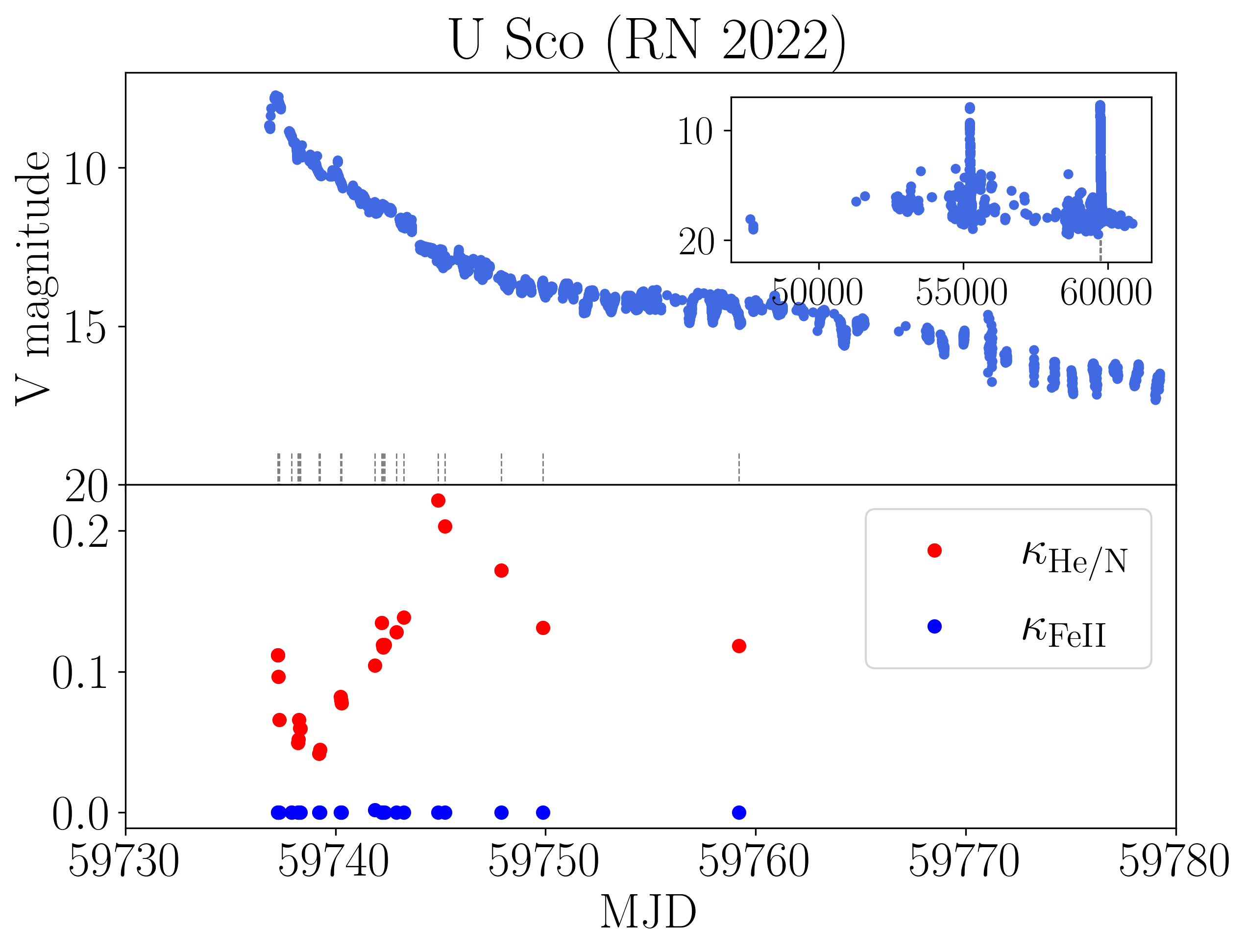}
    \includegraphics[width=0.6\linewidth]{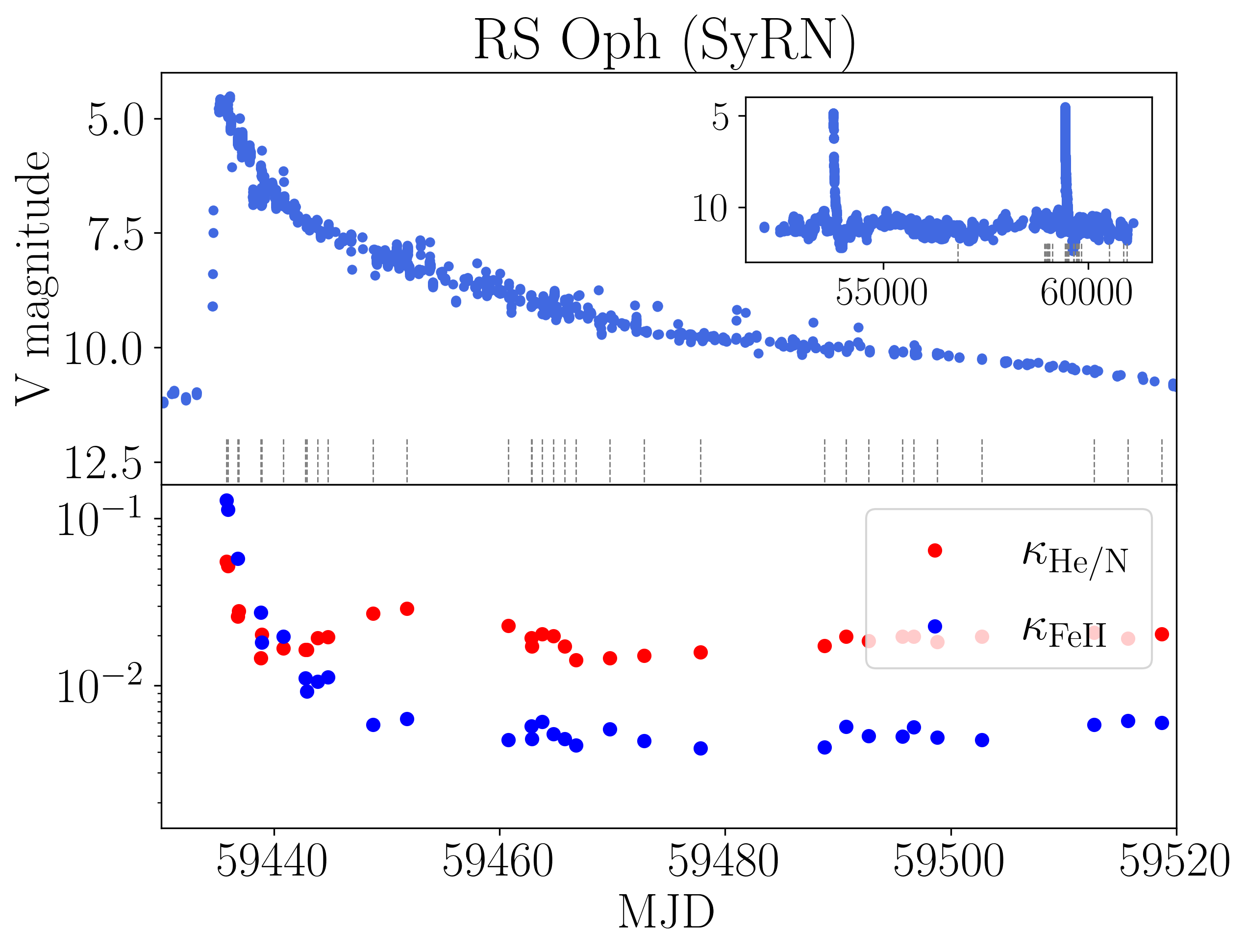}
\caption{
{\it Upper panel:} $V$ light-curves of the recurrent novae of our sample. {\it Lower panel:} Evolution of  $\kappa_{FeII}$ (blue) and $\kappa_{He/N}$ (red). 
The grey dashed lines indicate the epochs of the available spectra. 
Note that for RS\,Oph, the \ion{Fe}{2} flux in the first spectrum is null. 
Consequently, the corresponding blue point is not shown because the y-axis is on a logarithmic scale.}
\label{fig:rest_rec}
\end{figure}

Our sample includes two recurrent novae, namely U\,Sco and RS\,Oph. 
In both cases the spectra have been obtained around the most recent eruption (Fig.~\ref{fig:rest_rec}). 
In U\,Sco the first spectrum was obtained just right after the peak. 
No transition between phases is observed, since the \ion{Fe}{2} flux is almost zero everywhere. 
On the other hand, the first spectrum of RS\,Oph was obtained just before maximum. 
This is He/N-dominated, representing an early He/N phase, followed by an \ion{Fe}{2} phase and a subsequent late He/N phase associated with the long-term brightness fading.

\section{Discussion}

This work presents the investigation of the time evolution of the relative prevalence of \ion{Fe}{2} or He and N emission lines in the pre-nebular spectra of a sample of 26 novae using data from the ARAS Database.  
The prevalence of these lines is quantitatively derived by the $\kappa$ coefficient defined here. 
The time evolution of this coefficient is found to follow closely the nova light-curve, with 
$\kappa > 1$ (\ion{Fe}{2}-dominated) at early phases, when bright, and 
$\kappa < 1$ (He/N-dominated) in late phases, during the nova long-term fading, when faint. 
The spectral class is therefore a property that evolves with the nova brightness. 
This also occurs in novae with complex light curves exhibiting jitters (Fig.~\ref{fig:jitters}). 
During these brightness fluctuations, $\kappa$ repeatedly crosses unity, alternating between values above and below one. 
Consequently, the nova transitions back and forth between the \ion{Fe}{2}- and He/N-dominated spectral phases.

It is very remarkable that a significant fraction of the novae in the ARAS sample, 16 out of 26 (Figs.~\ref{fig:allfigs}, \ref{fig:jitters}, and \ref{fig:rest_rec}), shows transitions from \ion{Fe}{2}-dominated (defined as $\kappa > 1$) to He/N-dominated (defined as $\kappa < 1$) phases, and vice versa. 
Another nova (U\,Sco) has been noted to exhibit a transition from a He/N to an \ion{Fe}{2} phase \citep{2024MNRAS.527.9303A}. 
Among the novae in our sample that do not show spectral type transitions, six show trends toward a transition from one spectral type to the other, with the value of $\kappa$ approaching but not crossing unity.  
Five of these novae are classified as \ion{Fe}{2}, but their spectra were obtained during the bright phase and did not follow the brightness fading, when novae transition to the He/N spectral class. 
On the other hand, the rapidly evolving ($t_2$=1 d) Nova Her\,2021 only shows He/N-dominated spectra, most likely because no spectra were obtained during the short time interval of the brightness peak, but later, during the nova fading. 
It can thus be envisaged that the transition between spectral classes occurred in these six novae, but it was missed. 

The evolution between spectral types of novae was recently probed observationally and its origin discussed by \citet{2024MNRAS.527.9303A}.  
This work settled three different spectral phases before the nova enters the nebular phase: 
(1) an early He/N phase, on the rise towards the brightness peak, 
(2) an \ion{Fe}{2} phase, during the early decline in brightness, and 
(3) a late He/N phase, during the long-term brightness decline. 
The initial spectral phase dominated by \ion{He}{1}, \ion{N}{2}, and \ion{N}{3} emission lines finishes when the nova enters the so-called {\it iron curtain} phase. 
At this moment, the ejecta density is still sufficiently high, but its expansion decreases its temperature such that ultraviolet radiation is reprocessed to lower energies, leading to the emergence of optical \ion{Fe}{2} lines. 
As the ejecta continue expanding, the density decreases, making the plasma optically thin and allowing the inner, more highly ionized regions to become visible. 
At this stage, strong lines of \ion{He}{2}, \ion{N}{2}, and \ion{N}{3} appear \citep{2022ApJ...939....1H}.

The careful spectral sampling at different phases of the nova light-curve allowed \citet{2024MNRAS.527.9303A} to robustly establish the transitions among different spectral classes in ten sources. 
Very particularly, the acquisition of spectra on the rise made clear the existence of an early He/N-dominated phase. 
The work presented here, based on spectra drawn from the ARAS Database,  does not cover such early stage of nova evolution, but for RS\,Oph. 
There are four sources in common with \citet{2024MNRAS.527.9303A}, namely novae Del\,2013, Cas\,2021, Vul\,2021, and U\,Sco.  
Nova Del\,2013 was qualitatively found to evolve from the \ion{Fe}{2} to the He/N phase $\sim28$ d after t$_\mathrm{max}$, whereas here this transition is quantitatively computed based on $\kappa$ to have occurred $\sim37$ d after t$_\mathrm{max}$. 
Although the quantitative and qualitative criteria might differ, we note the higher time coverage of the spectra presented here. 
Meanwhile the frequent transitions from one phase to another of novae Cas\,2021 and Vul\,2021 reported by \citet{2024MNRAS.527.9303A} are confirmed here with values of $\kappa$ crossing the unity value (in both directions) up to eleven times in the case of Cas\,2021. 
This oscillatory behavior follows exactly the jitters in its light-curve, but we note that the initial data point in Nova Vul\,2021, with a value of $\kappa <1$ is assigned to the early He/N phase by \citet{2024MNRAS.527.9303A}. 
Finally, U\,Sco is found to transit from the early He/N phase to the \ion{Fe}{2} phase, but this transition is missed by our time coverage. 
To sum up, the present work increases the sample of sources presenting transitions among different spectral classes from 10 to 22 out of a sample of 32 novae. 
The transition thus occurs for $\simeq70$\%, thus it seems that the transition between spectral types is a quite common (may be even universal) behavior of novae before they enter into the nebular phase.

This evolutionary sequence is not monotonic, but there are sources that transit forth and back among spectral classes (as shown in Fig.~\ref{fig:jitters}). 
Such an example was presented by \citet{2024MNRAS.527.9303A}, which showed that V1405\,Cas transited at least three times from an \ion{Fe}{2} to a He/N phase. 
These transitions follow jitters in the light-curve, implying recurrent changes in the optical depth and ionization state of the ejecta that also affect the nova brightness.  
This is also the behavior found here for novae Cyg\,2014, Cas\,2020, and Vul\,2021 (Fig.~\ref{fig:jitters}).

\begin{figure}
    \centering
    \includegraphics[width=0.7\textwidth]{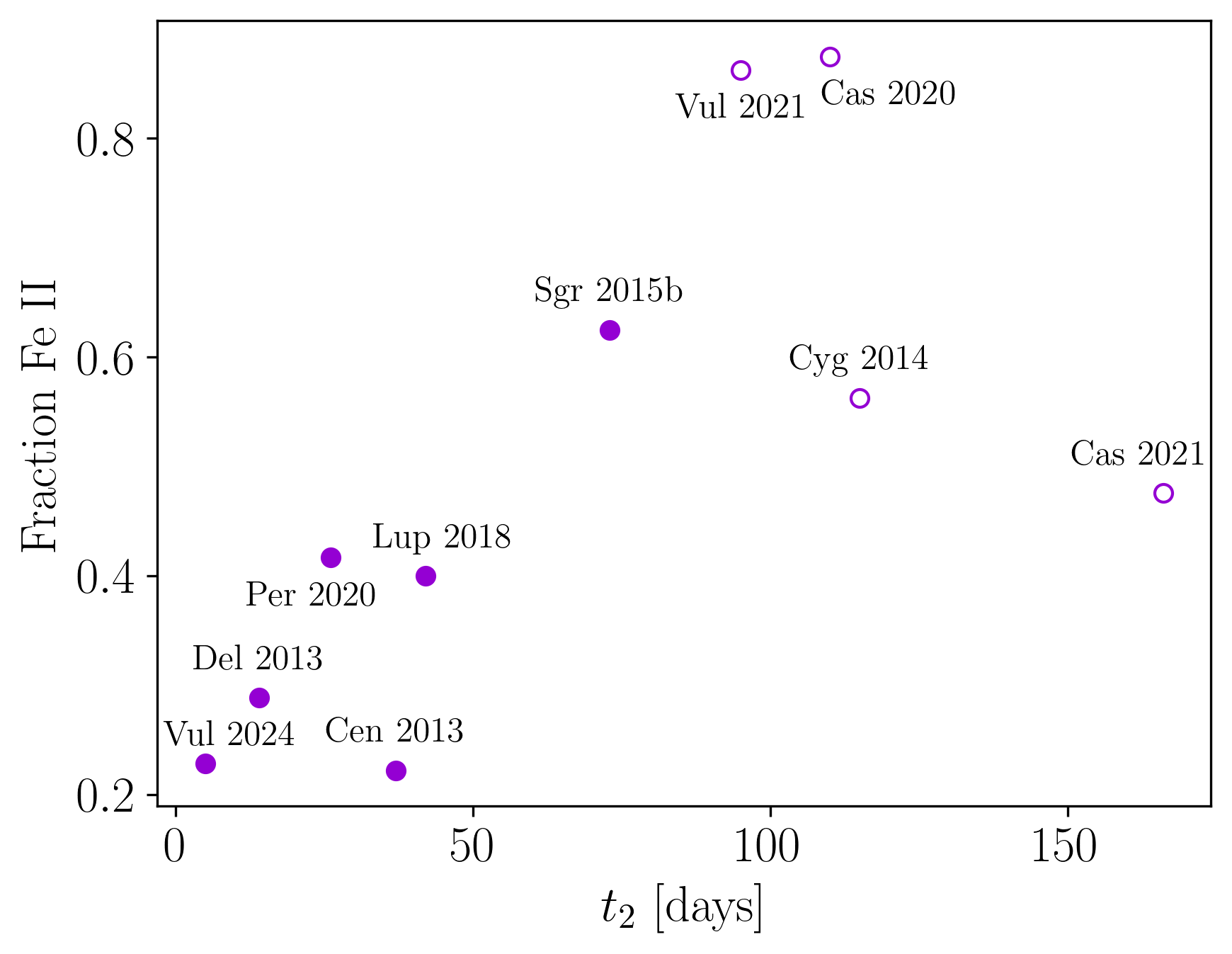}
    \caption{Fraction of the available spectra of the novae that show evolution from  the \ion{Fe}{2}-dominated to the He/N-dominated phase as a function of $t_2$, when available. 
    Open circles correspond to novae that exhibit jitters in their light curves.
    }
    \label{fig:frac_t2}%
    \end{figure}

The results of this study further confirm the idea that the \ion{Fe}{2} and He/N spectral classes are associated with different evolutionary phases of a nova
\citep{2012AJ....144...98W,2012BASI...40..185S,2014ASPC..490..145S,2024MNRAS.527.9303A}. 
Since the \ion{Fe}{2} phase occurs at the light-curve peak, whereas the He/N phases correspond to the early rise and later decline, the assignment of a spectral type to a nova can be expected to have severe biases associated with its speed class. 
Novae spending a long time period at peak in systems associated with older stellar populations should be easily observed in the \ion{Fe}{2} phase, whereas faster eruptions in younger populations with higher ejecta expansion velocities and WD mass are most easily observed in the late He/N phase.  
In order to examine this bias, we present in Figure~\ref{fig:frac_t2} the fraction of the spectra for each nova that is \ion{Fe}{2}-dominated as a function of $t_2$. 
This plot shows a broad positive correlation between the fraction of \ion{Fe}{2} spectra and $t_2$, such that the fraction of \ion{Fe}{2} spectra increases with $t_2$. 
This correlation supports the argument that it is more likely to `catch' the \ion{Fe}{2}-dominated phase when the nova evolution is slow. 
 
We note that $t_2$ is indeed related to the physical properties of the nova system, in particular the WD mass. 
Thus, this correlation does not imply that the observed spectral classification is independent of the underlying physical properties of the nova. 
Rather it highlights an additional observational bias arising from the different durations of the spectroscopic phases in novae with different progenitors.

Moreover, Table \ref{tab:novae} shows that the only classical nova classified as He/N is the very fast ($t_2 = 1$ d) Nova Her\,2021, whereas the four novae classified as \ion{Fe}{2} with known values of $t_2$, namely novae Sgr\,2016b, Sgr\,2016c, Sgr\,2021a and Sct\,2018, have longer values of $t_2$. 
The decline with time of $\kappa$ for the first three of these novae suggests that later spectra would have shown a transition to the He/N phase. 
Furthermore this observational bias favors the classification of a nova as \ion{Fe}{2}, as spectra are obtained typically a short time period after the peak, while the acquisition of spectra in later phases, when the nova is much fainter, becomes difficult or unfeasible.

The upper-right part of the diagram in Fig.~\ref{fig:frac_t2} is occupied by four novae whose light-curves show an extended period of high brightness superposed by large-amplitude jitters (open circles), as it is likely to observe a \ion{Fe}{2} phase because they stay bright for long-time periods.
One such case is Nova Cas\,2021 \citep{2024AstL...50..317T,2026NatAs..10..271A} whose frequent transitions between the \ion{Fe}{2} and He/N spectral classes described here and by \citet{2024MNRAS.527.9303A} are associated with the jitters superposed on a plateau phase of its light-curve. 
In the past, objects that show both \ion{Fe}{2} and He/N spectra at different epochs were classed as ``hybrid''.  
Indeed these four novae are classified as hybrids. 
Very notably, other novae classed as hybrid in the literature also present jitters in their light-curves, as V5558\,Sgr \citep{2011PASJ...63..911T} and V458\,Vul (Nova Vul\,2007; \citealt{2008Ap&SS.315...79P}; \citealt{2012ApJ...755..158R}), and the recurrent nova T\,Pyx (e.g. \citealt{2010ApJ...708..381S,2011A&A...533L...8S,2013ApJ...773...55S,2014ApJ...788..130C,2014AJ....147..107S,2024A&A...686A..72I,2026ApJ..1000..219S}), which shows similar photometric behavior near peak. 
These results suggest an additional observational bias in nova spectral classification. 
Because novae with a light-curve characterized by a plateau-like phase with jitters remain relatively bright and experience repeated transitions between the \ion{Fe}{2} and He/N phases associated with jitters, spectroscopic monitoring is more likely to capture these spectral transitions. Consequently, objects with these light-curve properties are commonly classified as ``hybrid''.


\section{Conclusions}

In this study, we used nova spectra from the ARAS Database to investigate the spectral evolution of novae. 
The spectral class of a nova at a given epoch is quantitatively characterized using the coefficient $\kappa$, defined as the ratio between the flux of the \ion{Fe}{2} $\lambda5169$ emission line and the mean flux of five representative He and N emission lines. 
Spectra with $\kappa > 1$ are assigned to the \ion{Fe}{2} phase, while those with $\kappa < 1$ to the He/N phase.

\noindent Our results are the following:
\begin{enumerate}

\item 
A large fraction of the novae in our sample exhibit transitions between the \ion{Fe}{2}- and He/N-dominated phases. 
Fifteen of the 24 classical novae and one of the two recurrent novae show at least one transition across $\kappa = 1$, with another one (U\,Sco) known to exhibit this transition in spectra before the brightness peak. 
Accounting for the novae reported by \citet{2024MNRAS.527.9303A}, the fraction of novae with spectral transitions is close to 70\%.
Among the remaining novae, six show trends towards such transition, although the time coverage of the available spectra was insufficient to cover it. 
A better time coverage of the spectroscopic data would have probably increased the sample of novae with spectral transitions. 
    
\item 
The coefficient $\kappa$ closely follows the optical light-curve, indicating that the spectral class is associated with the nova brightness variations.
Very particularly, $\kappa$ crosses unity multiple times when novae exhibit jitters, revealing repeated transitions between the \ion{Fe}{2}- and He/N-dominated phases. Therefore the spectral evolution is not always monotonic, but can be temporarily reversed, likely reflecting fluctuations in the optical depth of the ejecta.

\item 
Our quantitative analysis is in very good agreement with previous qualitative studies of individual novae,  particularly novae Del\,2013,  Cas\,2021, and Vul\,2021 \citep{2024MNRAS.527.9303A}. Different temporal sampling available between the studies provides a slightly different determination of the transition epochs. 
In both works it is observed that repeated oscillations are associated with light-curve jitters. 
The coefficient $\kappa$ defined here is therefore a reliable tool to describe the spectral evolution of novae. 

\item 
There is a bias favoring the classification of slow novae as \ion{Fe}{2} because their evolution on longer time-scales makes more likely capturing spectra during this phase. 
There is also a bias favoring the classification of nova with light-curves showing a plateau and jitters as hybrid, given their frequent transitions between the two spectral phases associated with jitters. 

\end{enumerate}

Our results provide further evidence that the traditional \ion{Fe}{2} and He/N spectral classifications represent an evolutionary sequence after a nova event including an early He/N phase during brightness rise, an \ion{Fe}{2} phase at peak, and a late He/N phase during the brightness decline. 

%
Since novae are usually spectroscopically classified around maximum light, this results in a strong observational bias associated with their brightness and photometric evolution. 
Thus faint, slow  novae are likely to be ``caught'' during the \ion{Fe}{2}-dominated phase, as the inability to obtain spectra once the nova has faded prevents detecting their subsequent evolution toward the He/N phase.
On the other hand, the \ion{Fe}{2} phase is easily missed in rapidly evolving novae, leading to a He/N classification. 

Since the time evolution of a nova is determined by the progenitor system intrinsic properties, particularly the WD mass, they also regulate the time-scale, duration, and relative prominence of the \ion{Fe}{2} and He/N phases and therefore the observational bias leading to their classification based on early and sparse spectroscopic information.

%
Finally, we note that novae that exhibit jitters in their light-curves, with multiple brightness peaks and minima, are commonly classified as hybrid.

The correlation between spectral and light-curve classification is not unambiguous. 
Indeed a simple photometric classification does not seems sufficient to ascertain their time evolution. 
For instance, novae Cen\,2013 and Cas\,2020 have D-type light-curves, but their light-curves are very different. 
The former declines very fast with a subtle dip (S+D), whereas the latter shows a plateau with multiple jitters and a large dip (J+D). 
Moreover the determination of time-scales ($t_2$ or $t_3$) of novae with jitters does not really follow the same criterion as for S-type light-curves. 
It seems that a more comprehensive description of the nova light-curves (and time-scales determination) is required to further assess their link with the intrinsic properties of the progenitor system.



\begin{acknowledgments}
 We thank the anonymous referee for the useful comments that helped to improv the clarity of the paper. M.K.\ acknowledges financial support from the Severo Ochoa grant CEX2021-001131-S funded by MCIN/AEI/10.13039/501100011033 through the "Incoming Visit" program and from AGAUR, CSIC, MCIN and AEI 10.13039/501100011033 under project PIE 20215AT016, Unidad de Excelencia María de Maeztu CEX2020-001058-M.
M.A.G.\ acknowledges financial support from the Severo Ochoa grant CEX2021-001131-S funded by MCIN/AEI/10.13039/501100011033 and the MCIU grant PID2024-162229NB-I00 cofunded with FEDER funds. 
DAVT thanks the support from UNAM DGAPA PAPIIT project IN102324 and acknowledges the financial support of the SECIHTI doctoral fellowship (CVU No. 1147167).
We gratefully acknowledge the invaluable contribution of the amateur astronomers who generously made their spectra available through the \textit{ARAS Database}. The number in parentheses indicates the number of spectra from each observer used in this work. In particular, we thank Pavol A. Dubovsky (258; Kolonica Observatory, SK), Terry Bohlsen (147; Mirranook Observatory, Armidale, AU), Forrest Sims (125; Desert Celestial Observatory, US), David Boyd (64; West Challow Observatory, UK), Christian Buil (30; Castanet Observatory, FR), David Cejudo (28; El Gallinero Observatory, ES), Paul Luckas (13; Shenton Park Observatory, AU), Tomas Medulka (12; Kolonica Observatory, SK), Paolo Berardi (8; Bellavista Observatory, IT), M.~Rodriguez (8; Madrid-Ventilla Observatory, ES), Tim Lester (7; Arnprior Observatory, CA), Jacques Montier (7; C.A.L.C. Observatory, FR), Kevin Gurney (7; KGH Observatory, UK), François Teyssier (6; Rouen Observatory, FR, and Sta.~Maria de Montmagastrell Observatory, ES), Peter Somogyi (4; TAT Observatory, HU), Bertrand Guegan (4; La Montagne Observatory, FR), Yana Markus (3; Kolonica Observatory, SK), Christophe Boussin (2; OCT Observatory, FR), J.~Hinnefeld (1; IUSB Observatory, US), Robert Barsa (1; Kolonica Observatory, SK), A.~Garcia (1; Sta.~Maria Observatory, ES), and Umberto Sollecchia (1; L'Aquila Observatory, IT). We acknowledge with thanks the variable star observations from the AAVSO International Database contributed by observers worldwide and used in this research.
\end{acknowledgments}





%
\facilities{\href{https://aras-database.github.io/database/index.html}{Astronomical Ring for Amateur Spectroscopy (ARAS)
Database} and,
\href{https://www.aavso.org}{American Association of Variable Stars Observers
(AAVSO).}}



\appendix

\section{Automated line flux determination}
\label{ap.1}

This work has finally considered 714 spectra of 26 nova.  
Such large dataset makes unfeasible to measure ``manually'' the flux of the \ion{Fe}{2}, He, and N emission lines of interest.  
Therefore an automated procedure to measure the flux of these emission lines has been developed.

As a first step, a continuum fit and subtraction has been performed to bring all spectra to a common zero level.
The continuum was estimated using an iterative spline fit with outlier rejection. We first select wavelength regions expected to trace the continuum and fit a smoothing spline (\texttt{UnivariateSpline}) to these points. The smoothing factor is scaled with the variance of the flux in the selected regions and the number of data points, so that the fit adapts to the noise level. After each fit, the residuals between the spectrum and the spline are computed, and points deviating by more than 2$\sigma$ are excluded from the continuum mask, thereby removing emission lines and other strong features from the fit. This procedure is repeated for four iterations, gradually refining the continuum estimate. The final spline model is taken as the continuum and subtracted from the original spectrum.

\begin{figure}
\centering
\includegraphics[width=0.5\textwidth]{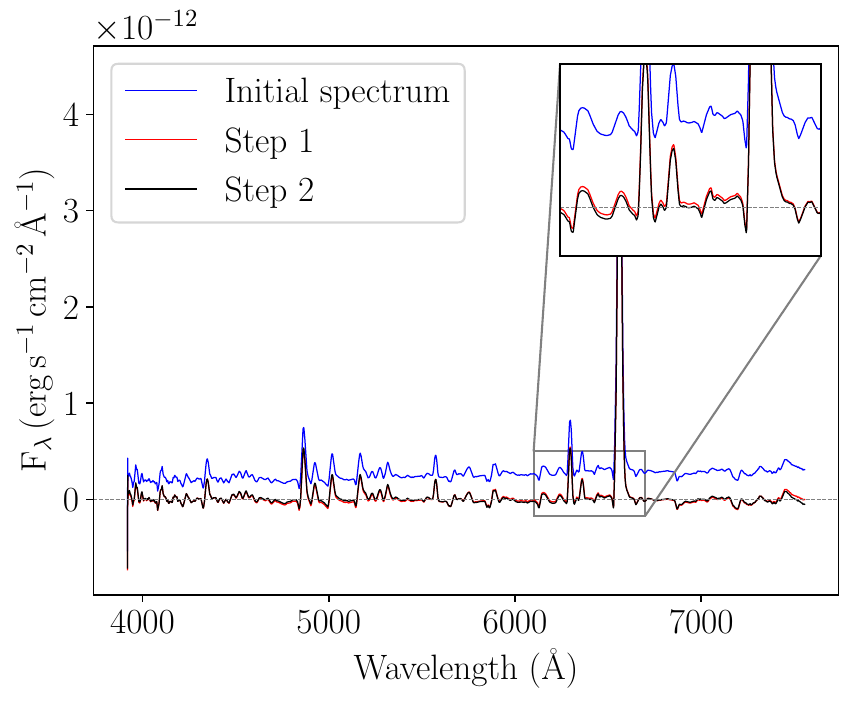}
\caption{
Example of continuum subtraction. 
The original spectrum is shown in blue, the first-order continuum-subtracted spectrum in red, and the second-order corrected spectrum is shown in black. 
The horizontal gray dashed line marks the zero-flux level.
}
\label{fig:example_spec_cont}%
\end{figure}

The initial procedure provides a good estimate of the continuum, but it must be noticed the complex spectral shape of nova spectra, with a large number of broad and in many cases blended emission lines, as well as absorption features. 
The continuum-subtraction was then further refined with a second, more restrictive iterative spline fit applied to the continuum-corrected spectrum. This step is designed to better capture local variations by focusing on points very close to the continuum level. At each iteration, we fit a higher-order smoothing spline (k=5) to the masked spectrum, with a smoothing factor scaled to the variance of the selected points. The continuum mask is then updated by retaining only those points with residuals within a narrow range around zero ($\pm 0.15\sigma$), thereby focusing on regions tightly clustered around the continuum and minimizing the contribution from emission lines or other features. This procedure is repeated for 10 iterations, progressively refining the continuum estimate. The final spline model is evaluated over the full wavelength range and subtracted from the original spectrum. An example of the continuum subtraction in presented in Fig. \ref{fig:example_spec_cont}.

\begin{figure}
\centering
\includegraphics[width=0.6\textwidth]{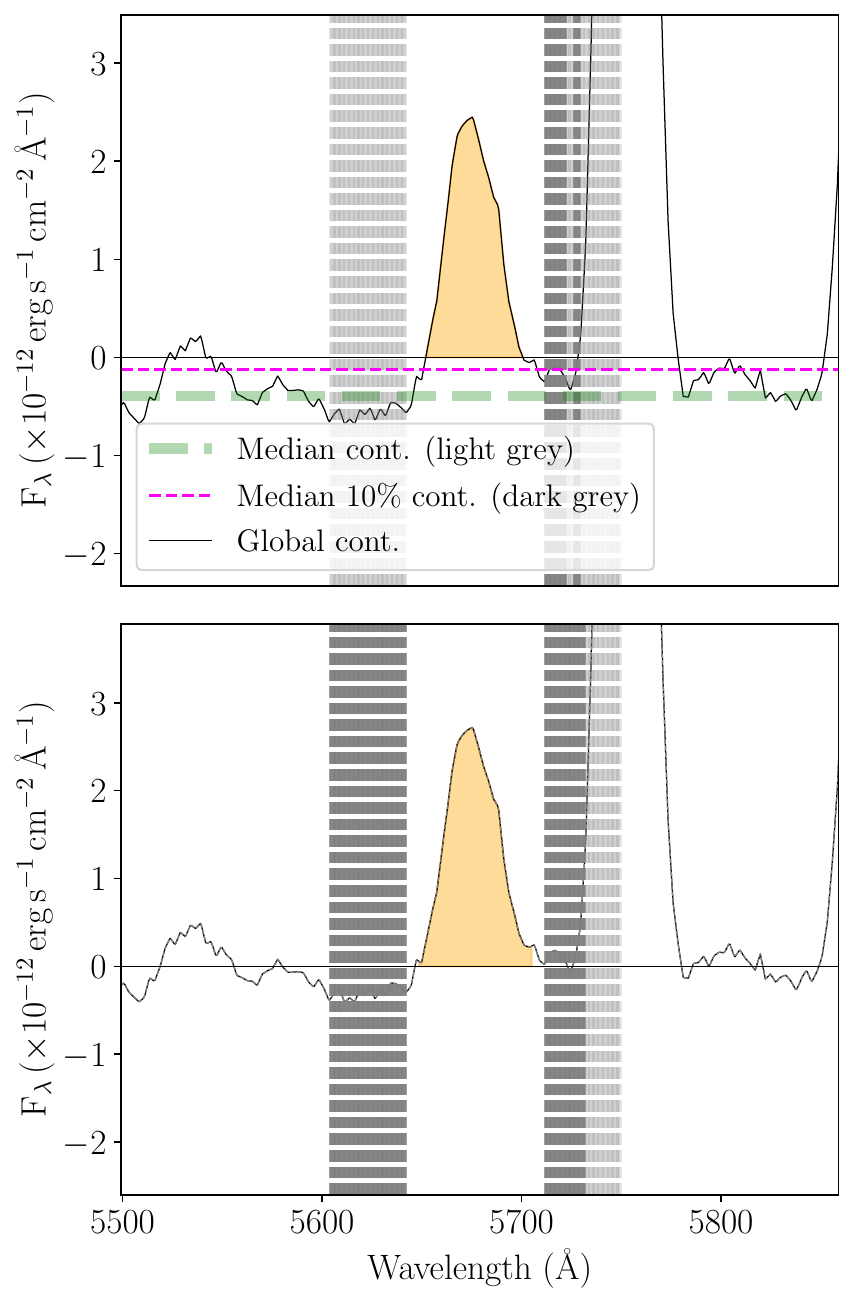}
\caption{
Example of the method followed for the determination of the local continuum and measurement of a line flux. 
{\it Top-panel:}
Spectral window of an emission line with global continuum (solid thin horizontal line at zero) and integrated flux (orange-shaded area). 
The thick-dashed green line shows the continuum calculated from the spectral apertures at the edge of the spectral window ($\pm80$ \AA\ around the rest-frame wavelength of the target line), as denoted by the light gray regions. 
The thin-dashed magenta line shows the continuum calculated from the edge regions but only for fluxes $\le$ 10\% of the peak flux of the emission line (dark gray). 
Since there is some difference between the two lines, we move the spectrum upwards with a shift corresponding to this difference. 
{\it Bottom-panel:}
Shifted spectrum with the final continuum determined from the median flux of the edge regions with fluxes $\le$ 50\%\ of the peak flux of the emission line (dark gray). 
This procedure ensure avoiding nearby emission or absorption features.
}
\label{fig:flux_meas}
\end{figure}

The emission-line fluxes were then measured using a local continuum subtraction followed by direct numerical integration of the line profile. 
The latter was preferred over a spectral fit because the line shape is generally not Gaussian. 
The local continuum is defined to determine the ``basal level'' from which the line flux will be measured, but also to calculate its signal-to-noise ratio (S/N). 
For each line, we extract a spectral window of $\pm 80$ \AA\ around the expected rest-frame wavelength, covering both the emission feature and nearby continuum regions.  
Two different values of the local continuum were then estimated: 
one from the median value of spectral apertures at the edges of this window to avoid the emission line (light gray region in the top panel of Fig.~\ref{fig:flux_meas}), and one more restrictive from the median value of these spectral apertures that have values within $\pm$10\% of the emission line peak (dark gray region in the top panel of Fig.~\ref{fig:flux_meas}). 
These two values of the continuum were subsequently compared and, if different, a new continuum level was computed from the median value of these spectral apertures with values within $\pm$50\% of the emission line peak (dark gray region in the bottom panel of Fig.~\ref{fig:flux_meas}). 
These aperture regions for the continuum calculation were enforced to cover at least 20\% of the edge regions to ensure that it is computed from a representative spectral sample. 
If this was not the case, the median value of the edge regions of the initial spectrum, considering only values within $\pm$50\% of the emission-line peak, was adopted as the continuum. If the aperture regions still did not cover at least 20\% of the edge region, the initial median flux of the edge regions was adopted (light gray region in the top panel of Fig.~\ref{fig:flux_meas}). 
An inspection of the continuum estimate showed that this procedure provides a stable estimate of the local continuum level even in noisy and complex spectra.

Once the integration range is defined, the line flux is obtained by trapezoidal integration over the continuum-subtracted spectrum. 

The S/N ratio is computed using the equation: 
\begin{equation}
    S/N = \frac{\sum_i^N F_{\it i}}{(rms\times\sqrt N)},
\end{equation}
where $F_i$ is the flux at each spectral element $i$ used for the line flux calculation, $rms$ is the mean square root of the continuum (computed as described above), and $N$ is the number of spectral elements used for the calculation of the flux of the emission line. 
Emission lines measurements with a S/N ratio lower than 10 were discarded for subsequent analysis. Moreover each measurement is assigned a quality flag when the continuum determination is uncertain or when the line profile appears affected by blending or contamination. 

\bibliography{novae}{}
\bibliographystyle{aasjournalv7}



\end{document}